\documentclass[aps,showpacs,superscriptaddress,nofootinbib,amsmath,amssymb,10pt]{revtex4}
\usepackage[utf8]{inputenc}
\usepackage{graphicx}
\usepackage{amsmath}
\usepackage{amssymb}
\usepackage{enumerate}
\usepackage{xcolor}
\graphicspath{./fig}
\begin{document}
  \title{Higher order perturbative and nonperturbative QCD corrections on the proton structure functions and parity violating electron asymmetry}
\author{F. Zaidi}
\affiliation{Department of Physics, Aligarh Muslim University, Aligarh - 202002, India}
\author{M. Sajjad Athar\footnote{Corresponding author: sajathar@gmail.com}}
\affiliation{Department of Physics, Aligarh Muslim University, Aligarh - 202002, India}
\author{S. K. Singh}
\affiliation{Department of Physics, Aligarh Muslim University, Aligarh - 202002, India}



\begin{abstract}
We study the nonperturbative and higher order perturbative corrections on the electromagnetic ($F_{1p,2p}^\gamma$) and 
electromagnetic-weak interference ($F_{1p,2p,3p}^{\gamma Z}$) structure functions and their impact on the parity violating electron asymmetry
in the deep inelastic scattering of longitudinally polarized electron off an unpolarized proton target. The numerical results for them
are presented by including the perturbative corrections beyond the leading order (LO) up to the next-next-to-leading-order (NNLO) 
and nonperturbative QCD corrections due to the target mass corrections (TMC) and the higher twist (HT: twist-4) effects.
We also present the numerical results for the electron beam spin asymmetry $A_{PV}^{(e)}(x,Q^2)$ corresponding to the JLab 
energies of 6 GeV, 12 GeV and 22 GeV and discuss the feasibility of determining the $d/u$ quark distribution ratio. The results obtained in this work may be
 useful for the analysis of future measurements at the Electron Ion Collider(EIC) in USA, and the Electron ion collider in China(EicC) aimed at
 studying parity violating effects in the deep inelastic scattering of polarized electrons from unpolarized proton targets.

\end{abstract}
\maketitle

\section{Introduction}
The study of the parity violating (PV) electron asymmetry in the deep inelastic scattering (DIS) of polarized electrons from unpolarized 
and polarized proton targets has played an important role in understanding the structure of the nucleon and its interactions in the electroweak sector. The 
first observation of the parity violating asymmetry $A_{PV}^{(e)}(x,Q^2)$ in DIS of polarized electrons from unpolarized proton and deuteron targets
was made at SLAC~\cite{Prescott:1978tm, Prescott:1979dh}. The PV asymmetry $A_{PV}^{(e)}(x,Q^2)$ is defined as:
\begin{equation}
 A_{PV}^{(e)}(x,Q^2)=\frac{\sigma_R(x,Q^2)-\sigma_L(x,Q^2)}{\sigma_R(x,Q^2)+\sigma_L(x,Q^2)},
\end{equation}
where $x$ is the Bjorken variable, $Q^2$ is the four momentum transfer square, and $\sigma_{R,L}(x,Q^2)$ are the cross sections for the right (left)
handed polarized high energy electrons scattering from unpolarized protons. The experimental observations of $A_{PV}^{(e)}(x,Q^2)$ confirmed the
existence of the neutral currents (NC) in the weak sector of the electron-nucleon interactions and also determined the value of the weak mixing angle $\theta_W$ which was consistent with the 
predictions of the unified theory of weak and electromagnetic interactions proposed by Glashow, Salam and Weinberg (GSW)~\cite{Glashow:1959wxa, Salam:1959zz, Weinberg:1967tq}.
Since then several efforts have been undertaken to measure the parity violating asymmetry in DIS of polarized electrons.
These experiments played a major role in 
establishing the GSW model as the standard model (SM) of particle physics unifying the weak and electromagnetic interactions and provided
great impetus to perform further experiments on the measurements of the parity violating asymmetry in scattering of polarized electrons
and muons from nucleons and nuclear targets corresponding to the kinematic region of 
quasielastic and inelastic scattering in addition to the deep inelastic scattering (DIS)~\cite{SAMPLE:1999mku, SAMPLE:2003chl, SAMPLE:2003wwa, HAPPEX:2004lbc, HAPPEX:2006oqy, HAPPEX:2011xlw, G0:2005chy, G0:2009wvv, A4:2004gdl, 
Maas:2004dh, Baunack:2009gy, SLACE158:2005uay, Qweak:2014xey, Abrahamyan:2012gp, Becker:2013fya, MOLLER:2014iki}. Their theoretical analyses using the methods of perturbative QCD and lattice gauge theory
simulations of QCD~\cite{Cahn:1977uu, Bjorken:1978ry, Wolfenstein:1978rr, Fajfer:1984um, Castorina:1985uw, Kaplan:1988ku, Mckeown:1989ir, 
Hobbs:2008mm, Mantry:2010ki, Brady:2011uy, Hall:2013hta, Zhao:2016rfu, Bacchetta:2023hlw, Du:2024sjt} have led to important information about
understanding the nucleon structure in QCD.

In the specific case of the deep inelastic scattering of polarized electrons from unpolarized protons and nuclei, the various experiments measure the
spin dependent cross sections $\sigma_R(x,Q^2)$ and $\sigma_L(x,Q^2)$ corresponding to the scattering of right handed $(R)$ and left handed ($L$)
polarized electrons from protons which determine the parity violating electron asymmetry $A_{PV}^{(e)}(x,Q^2)$ and depend on the proton structure 
functions $F_{1p,2p,3p}^{\gamma,\gamma Z,Z}(x,Q^2)$ and the weak neutral current electron couplings $g_V^e$ and $g_A^e$ in the vector ($V$) and axial
vector $(AV)$ interactions. Here $F_{1p,2p,3p}^{\gamma(Z)}(x,Q^2)$ ($F_{3p}^{\gamma}(x,Q^2)=0$) are the proton structure functions due to $\gamma$, and 
$Z$ interactions shown in Fig.~\ref{feyn0} while $F_{1p,2p,3p}^{\gamma Z}(x,Q^2)$ are the proton structure functions arising due to 
the interference of $\gamma$($Z$) interactions. In all the terms appearing in the cross section the contribution of the $\gamma Z$ ($Z$) structure functions are 
suppressed by a factor $\eta^{\gamma Z} \big((\eta^{\gamma Z})^2\big)$, where $\eta^{\gamma Z}$ is given by
\begin{equation}\label{coup}
\eta^{\gamma Z}=\frac{G_F M_Z^2 Q^2}{2\sqrt{2} \pi \alpha (Q^2+M_Z^2)},
\end{equation}
with $G_F=1.16 \times 10^{-5}$ GeV$^{-2}$ is the Fermi coupling constant, $\alpha$ is the fine structure constant
 and $M_Z=91.1876\pm 0.0021$ GeV is the mass of $Z-$boson.
In general, the value of $\eta^{\gamma Z}$ for $Q^2<< M_Z^2$ is quite small, i.e., of the order of $10^{-3}$ for $Q^2\simeq10$ GeV$^2$. Therefore, the contributions 
of the $\gamma Z$ and $Z$ structure functions $F_{1p,2p,3p}^{\gamma Z, Z}(x,Q^2)$ to the cross sections $\sigma_R(x,Q^2)$ and $\sigma_L(x,Q^2)$
are small and negligible as compared to the contributions of the $\gamma$ structure functions $F_{1,2}^\gamma(x,Q^2)$. However, the parity violating electron asymmetry
$A_{PV}^{(e)}(x,Q^2)$ which is defined in terms of the difference of the helicity dependent cross sections $\sigma_R(x,Q^2)$ and $\sigma_L(x,Q^2)$, depends 
on the $\gamma Z$ structure functions $F_{1p,2p,3p}^{\gamma Z}(x,Q^2)$ and on the weak neutral current couplings of the electron. 
The $\gamma Z$ structure functions $F_{1p,2p,3p}^{\gamma Z}(x,Q^2)$ depend also upon the neutral current couplings of the quarks 
involving the weak mixing angle $\theta_W$ and the parton distribution functions (PDFs) in the nucleon. Therefore, a precise measurement of the parity
violating electron asymmetry can be used to extract the weak mixing angle $\theta_W$ and to study the quark PDFs in nucleons. Knowledge of the quark PDFs is required 
to extract the value of the weak mixing angle $\theta_W$. However, the uncertainties associated with the determination of the quark 
PDFs, as discussed by Fu et al.~\cite{Fu:2025anw}, Bodek et al.~\cite{Bodek:2025kxl}, 
Lin~\cite{Lin:2025hka}, Gao et al.~\cite{Gao:2025hlm}, Chiefa et al.~\cite{Chiefa:2025cap}, Risse et al.~\cite{Risse:2025qlo} and many 
others in the past~\cite{Brock:2000ud, Giele:2001mr, Bozzi:2011ww, Lin:2017snn, Fu:2020mxl, Amoroso:2022eow},
affect the extraction of the mixing angle $\theta_W$. In an analysis of the uncertainties made by Amoroso et al.~\cite{Amoroso:2022eow}, it was
found that the uncertainties among different PDFs like NNPDF4.0~\cite{NNPDF:2021njg}, ATLASpdf21~\cite{ATLAS:2021vod}, MSHT20~\cite{Bailey:2020ooq}, CT18~\cite{Hou:2019efy}, ABMP16~\cite{Alekhin:2017kpj}, etc., are possibly a consequence of both 
adopted methodology and differences in data included. However, there is an overall agreement among the different PDFs sets in the kinematic 
regions of $x$ and $Q^2$, where experimental data are abundant or over constrained, such as in the intermediate region of $x$ specially in the 
case of light flavor quarks leading to a consensus on the extraction of the weak mixing angle $\theta_W$. Nonetheless, discrepancies
are observed in areas with scarce data or complex nuclear effects, such as for the region of small $x$ or large $x$ or for the heavy-flavor PDFs.
Moreover, a simplification arises in the case of isoscalar targets, like the deuterium in which the PV electron asymmetry $A_{PV}^{(e)}(x,Q^2)$
calculated in the quark-parton model using the leading order (LO) of perturbative QCD valid in the kinematics of Bjorken scaling becomes independent of
the structure functions in the limit of isospin symmetry and depends only on the weak mixing angle $\theta_W$~\cite{Hobbs:2008mm}. This has been used to extract the weak mixing angle 
$\theta_W$ from the measurements of PV electron asymmetry in early experiments on the DIS of polarized electrons from 
deuteron targets assuming isospin symmetry~\cite{Prescott:1979dh, Hwang:1981rq, Heil:1989dz}. However, the nuclear effects due to the use of deuteron 
target and possible effects due to the violation of isospin symmetry in deuteron, though small should be estimated. Alternatively, it has been suggested 
in literature to study the effects of the violation of isospin symmetry in the quark PDFs using the parity violating electron asymmetry measurements from the DIS
of polarized electrons from deuteron targets~\cite{Prescott:1979dh, Hwang:1981rq, Heil:1989dz, Cocuzza:2026vey}.

The importance of the PV electron asymmetry $A_{PV}^{(e)}(x,Q^2)$ in extracting the weak mixing angle and its anticipated role in studying the isospin
symmetry violation of quark PDFs has initiated many experimental proposals to make precise measurements of $A_{PV}^{(e)}(x,Q^2)$ with high energy and high intensity polarized 
electrons beams at various electron accelerators. For example, the SoLID experiment and its upgrade at JLab~\cite{JeffersonLabSoLID:2022iod, Accardi:2023chb, Meziani:2024leh, Cotton:2024rpn}
in USA, at HIAF~\cite{Anderle:2021wcy} in China, and at BNL~\cite{AbdulKhalek:2021gbh, Boughezal:2022pmb, Nadel-Turonski:2025sfv} in USA are at various
stages of development to make these measurements. These experiments will measure the helicity dependent cross sections $\sigma_R(x,Q^2)$ and $\sigma_L(x,Q^2)$ 
and determine the various electroweak nucleon structure functions $F_{1N,2N}^{\gamma}(x,Q^2)$ and $F_{1N,2N,3N}^{\gamma Z}(x,Q^2)$ and PV electron asymmetry with
improved precision in the DIS of polarized electrons from protons and deuterons.

In view of these experimental proposals various theoretical calculations have been recently made for the neutral current nucleon structure functions 
and the PV electron asymmetry $A_{PV}^{(e)}(x,Q^2)$. Notwithstanding the importance of the $A_{PV}^{(e)}(x,Q^2)$ in extracting the 
weak mixing angle which is straightforward in the region of the kinematics of Bjorken scaling $Q^2\to \infty,~\nu\to\infty$ using the quark parton model, most of the 
experiments are done at high but finite $Q^2$ and $\nu$, where various corrections due to the violation of Bjorken scaling may play a role.
At finite $Q^2$, the corrections due to higher order perturbative QCD and the nonperturbative corrections due to the target mass correction (TMC) and higher twist (HT) are important in a
wide region of $x$. In addition, the electroweak structure functions get contribution from the sea quarks and gluons in the region of lower $x$ which has been 
studied in some detail~\cite{Gallo:2008yh, H1:2015ubc, Zhao:2016rfu, Bertone:2024dpm}. In the region of higher $x$, the electroweak structure functions get contribution from the resonance excitations and pion 
production processes which can be minimized by choosing proper limits on $Q^2$ and $W$, the center of mass (CM) energy in order to apply the 
phenomenology of DIS. Generally, a kinematic limit of $W\ge 2$ GeV and $Q^2\ge 1$ GeV$^2$ is chosen to ensure that a pure DIS formulation 
is applied to extract the nucleon structure functions and the electron asymmetry $A_{PV}^{(e)}(x,Q^2)$.
In this kinematic region of $W\ge 2$ GeV and $Q^2\ge 1$ GeV$^2$, it is important to study the corrections due to 
higher order perturbative QCD and nonperturbative effects. In past there have been some calculations of the higher order perturbative QCD and HT effects
at NLO~\cite{Castorina:1985uw, Mantry:2010ki, Brady:2011uy, Hall:2013hta, Borsa:2022irn}, but the question of TMC has been addressed only recently in the calculations of the electroweak structure functions and 
the parity violating electron asymmetry~\cite{Brady:2011uy}.

In section~\ref{formalism}, we describe in brief the formalism for the DIS of polarized electrons from protons 
and derive an expression for the electron asymmetry
$A_{PV}^{(e)}(x,Q^2)$ in terms of the electromagnetic proton structure functions $F_{1p,2p}^{\gamma}(x,Q^2)$, 
and  $F_{1p,2p,3p}^{\gamma Z}(x,Q^2)$ the proton structure functions in the weak-electromagnetic interference sector.
We follow the works of Schienbein et al.~\cite{Schienbein:2007gr} to incorporate the TMC effect, and the works of Dasgupta et al.~\cite{Dasgupta:1996hh} 
and Stein et al.~\cite{Stein:1998wr} to include the higher twist effect up to twist-4. For the evolution of PDFs at NLO and NNLO,
 we follow the works of Vermaseren et al.~\cite{Vermaseren:2005qc} and Moch et al.~\cite{Moch:2004xu, Moch:2008fj} and use the parameterization 
 of the nucleon PDFs given by Martin, Motylinski, Harland-Lang, Thorne (MMHT)~\cite{Harland-Lang:2014zoa} in the MSbar scheme.
In section~\ref{results}, we present quantitatively the results for the effect of various corrections described above on these structure functions and 
discuss the validity of the Callan-Gross relation~\cite{Callan:1969uq} in the presence of these corrections in the $\gamma$ and $\gamma Z$ sectors.
For this purpose, we define the ratios $r_{p}^{\gamma /\gamma Z}(x,Q^2)=\frac{F_{2p}^{\gamma/ \gamma Z}(x,Q^2)}{2 x F_{1p}^{\gamma/ \gamma Z}(x,Q^2)}$ and 
$R_{p}^{\gamma /\gamma Z}(x,Q^2)=\frac{F_{Lp}^{\gamma/ \gamma Z}(x,Q^2)}{2 x F_{1p}^{\gamma/ \gamma Z}(x,Q^2)}$ and discuss their numerical results, where
$F_{Lp}^{\gamma/ \gamma Z}(x,Q^2)$ is the longitudinal structure function.
Furthermore, we present and discuss the quantitative results
for the single differential cross sections $\frac{d\sigma^{ep}}{dx}$, electron asymmetry $A_{PV}^{(e)}(x,Q^2)$ as well as the results for 
$a_{1p}(x,Q^2)=-g_A^e\frac{F_{1p}^{\gamma Z}(x,Q^2)}{F_{1p}^{\gamma}(x,Q^2)}$ (which may provide information about 
$d/u$ ratio) in presence of these corrections in the kinematic region of $x$ and $Q^2$ relevant for
future experiments planned at JLab~\cite{Arrington:2021alx, JeffersonLabSoLID:2022iod, Accardi:2023chb, Meziani:2024leh}, 
EIC~\cite{AbdulKhalek:2021gbh, Boughezal:2022pmb, Nadel-Turonski:2025sfv} and EicC~\cite{Anderle:2021wcy} to study the 
parity violation in the DIS of polarized electrons from nucleons. In section~\ref{summ}, we summarize our findings and conclude the present work.

\section{Formalism}\label{formalism}
\subsection{Differential scattering cross section}
\begin{figure}
 \centering\includegraphics[height= 3.5 cm, width=8 cm]{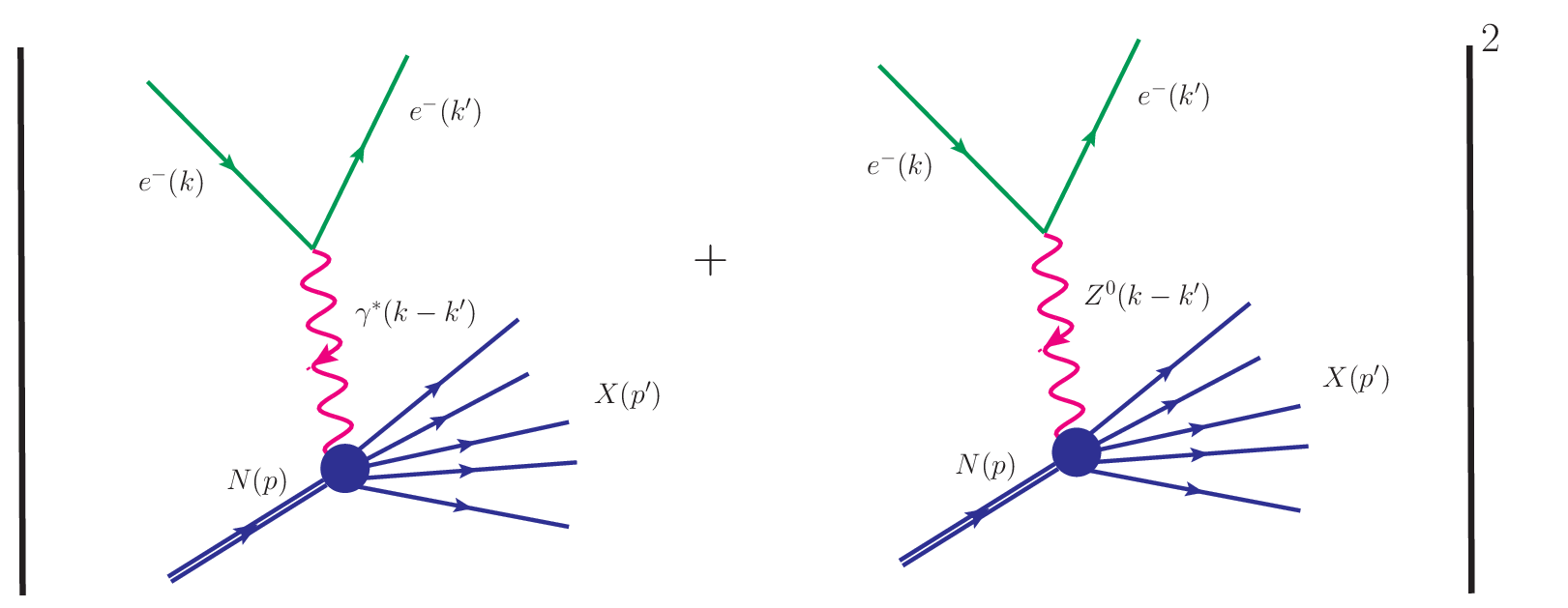}
 \caption{Diagrammatic representation of parity-violating contribution to the DIS which results from the interaction
of purely electromagnetic ($\gamma^\ast$ exchange) and purely weak ($Z$ exchange) amplitudes .}
\label{feyn0}
\end{figure}

The basic reaction for the DIS of polarized electron scattering off an unpolarized nucleon target is given by
 \begin{equation}\label{reac}
 \vec{e^-}(k,s_1)+p(p,s_p) \rightarrow  e^-(k^\prime,s_2)+X(p^\prime,s_X),
\end{equation}
where quantities within the parenthesis represent the four momenta and spin of the corresponding particles.
The general expression for the differential cross section in terms of the purely electromagnetic contribution via photon ($\gamma$) exchange
and the neutral current induced weak contribution via $Z-$exchange to the 
scattering amplitudes (as depicted in Fig.~\ref{feyn0}) is written as
\begin{equation}\label{xamp}
 \frac{d^2\sigma}{dx dy}=\frac{2\pi y \alpha^2}{q^4}\;\Big(|{\cal M_{\gamma}}+ {\cal M_{Z}}|^2 \Big),
\end{equation}
where $q^2=(k-k')^2$ is the four momentum transfer square,
$y=\frac{p\cdot q}{p \cdot k}=\frac{\nu}{E}$ is the inelasticity, $E$ is the incoming electron beam energy, $\nu=E-E'$ is the energy transfer and 
$E'$ is the energy of the scattered electron beam.
The differential cross section defined in Eq.~\ref{xamp} is expressed in terms of the leptonic tensors ($L^{\mu\nu,i}$) and 
the hadronic tensors ($W_{\mu\nu}^{i}$), where $i=\gamma,~Z,~\gamma Z$ correspond to the pure photon ($\gamma$) exchange, $Z$-exchange and 
their interference term $\gamma Z$, and is written as~\cite{Du:2024sjt, Boughezal:2022pmb}:
\begin{eqnarray}\label{dxsec0}
 \frac{d^2\sigma}{dx dy}&=&\frac{2\pi y \alpha^2}{q^4}\;\Big( \;L^{\mu\nu,\gamma}\;W_{\mu\nu}^{\gamma } \;+\;
 \eta^{\gamma Z} \;L^{\mu\nu,\gamma Z}\;W_{\mu\nu}^{\gamma Z} \;+\;\Big(\eta^{\gamma Z}\Big)^2 \;L^{\mu\nu, Z}\;W_{\mu\nu}^{Z} \Big)
\end{eqnarray}
The leptonic tensors for the aforementioned interaction channels are given by~\cite{ParticleDataGroup:2024cfk}:
\begin{eqnarray}
 L^{\mu\nu,\gamma}=2(k^\mu k'^\nu+k^\nu k'^\mu-k \cdot k' g^{\mu\nu}-i\lambda_e \epsilon^{\mu\nu\alpha\beta}k'_{\alpha} k_{\beta}),
 \label{lep_gammap}
\end{eqnarray}

\begin{eqnarray}
L^{\mu\nu,Z}&=&(g_V^e-\lambda_e g_A^e)^2\; L^{\mu\nu,\gamma}(\lambda_e),~~\textrm{and}
 \label{lep_z0}
\end{eqnarray}

\begin{eqnarray}
L^{\mu\nu,\gamma Z}(\lambda_e)&=&(g_V^e-\lambda_e g_A^e)\; L^{\mu\nu,\gamma}(\lambda_e)
 \label{lep_gammaz}
\end{eqnarray}
with $\lambda_e (=\pm 1)$ as the helicity of incoming electron,  
\begin{equation}\label{coups}
 g_V^e=-1/2+2 \\sin^2\theta_W\;\;\;\textrm{and}\;\;\;g_A^e=-1/2
\end{equation}
are the vector and the axial-vector couplings for the electron and $\theta_W$ is the Weinberg angle. 
If we choose the direction of the polarized electron beam along the z-axis and ignore its mass ($m_l \to 0$) then the four momenta are defined 
in the rest frame of nucleon as:
\begin{eqnarray}
 k^\mu&=&(k^0,\vec k)=(E,0,0,|\vec k|)=(E,0,0,E)\nonumber\\
 k'^\mu&=&(E',\vec k')=(E',E' \sin\theta \cos\varphi,E'\sin\theta \sin\varphi,E' \cos\theta)\nonumber\\
 q^\mu&=&k^\mu-k'^\mu\nonumber\\
q^0&=&E-E';\;\; q^x=-E' \sin\theta \cos\varphi;\;\;q^y=-E'\sin\theta \sin\varphi;\;\;q^z=E-E'\cos\theta\nonumber\\
p^\mu&=&(p^0,\vec p)=(M,\vec{0}),
\end{eqnarray}
where $\varphi$ is the azimuthal angle, and $\theta$ is the lab scattering angle.
The Bjorken variable $x$ is given by
\begin{equation}
 x=\frac{Q^2}{2 p \cdot q}=\frac{Q^2}{2 M \nu},
\end{equation}
with $M$ as the target nucleon mass.

The hadronic tensor ($W^j_{\mu\nu};~(j=\gamma,\gamma Z, Z)$) is generally defined in terms of the dimensionless nucleon structure 
functions $F_{ip}(x,Q^2)~;(i=1-5)$ with metric tensor $g_{\mu\nu}$, four momentum $p$ and the four momentum transfer $q$, as~\cite{ParticleDataGroup:2024cfk}:
\begin{eqnarray}\label{had_ten}
 W^j_{\mu\nu} &=& \Big(-g^{\mu\nu}+\frac{q_\mu q_\nu}{q^2}\Big)\;F^j_{1p}(x,Q^2) + \Big(p_\mu-\frac{p\cdot q}{q^2}q_\mu \Big)\;\Big(p_\nu-\frac{p\cdot q}{q^2}q_\nu \Big)\;\frac{F^j_{2p}(x,Q^2)}{p\cdot q}\nonumber\\
 &-&i\epsilon^{\mu\nu\rho\sigma}\;\;\frac{q_\rho p_\sigma}{2 p \cdot q}\;F^j_{3p}(x,Q^2)+
\frac{q_{\mu} q_{\nu}}{p\cdot q} F^j_{4p}(x,Q^2) +
                (p_{\mu}q_{\nu} + p_{\nu}q_{\mu}) F^j_{5p}(x,Q^2).
\end{eqnarray}
 By using Eqs.~\ref{lep_gammap}-\ref{lep_gammaz} and \ref{had_ten} in Eq.~\ref{dxsec0}, the differential scattering cross section is obtained in 
 terms of the dimensionless nucleon structure functions. Neglecting the lepton mass terms ($m_l \to 0$), the contribution from the 
 terms with $F_{4p,5p}(x,Q^2)$  will be zero and the cross section is then written as:
 \begin{eqnarray}\label{d2nc_comp}
 \frac{d^2\sigma}{dx dy}&=& \big(\frac{2\pi y \alpha^2}{q^4}\big)\;\big(\frac{4 M E}{y}\big)\;\Big[ xy^2 \;F_{1p}(x,Q^2)+ \big( 1-y-\frac{M x y}{2 E}\big)\;F_{2p}(x,Q^2)\nonumber\\
&& +x y \big(1-\frac{y}{2}\big) \lambda_e \;F_{3p}(x,Q^2).\Big]
\end{eqnarray}
The dimensionless proton structure functions $F_{ip}(x,Q^2);~(i=1-3)$ having contribution from the three terms corresponding to 
$\gamma-$, $Z-$ exchanges and their interference $\gamma Z$, are written as~\cite{ParticleDataGroup:2024cfk}:
\begin{eqnarray}
\label{fnc}
 F_{ip}(x,Q^2)&=& F_{ip}^{ \gamma}(x,Q^2)-\eta^{\gamma Z}\;(g_V^e-\lambda_e\;g_A^e)\;F_{ip}^{\gamma Z}(x,Q^2)+(\eta^{\gamma Z})^2\;(g_V^e-\lambda_e\;g_A^e)^2\;F_{ip}^{Z}(x,Q^2),
  \end{eqnarray}
In the quark-parton model, these dimensionless nucleon structure functions are derived in terms of the 
unpolarized quark and antiquark distribution functions $q_i(x)$ and $\bar q_i(x)$, respectively in the 
kinematic region of Bjorken scaling, i.e., $Q^2\to \infty$, $\nu \to \infty$, $\frac{Q^2}{2 M \nu}=x$ in which they scale.
Consequently, they depend only on one kinematic variable, i.e., $x$ (known as Bjorken scaling variable or Bjorken-$x$) and are therefore, expressed
as a function of $x$~\cite{Callan:1969uq, Cahn:1977uu, Du:2024sjt}:
\begin{eqnarray}
\label{cg0}
F_{1p}^j(x)&=& \frac{1}{2 x}\;F_{2p}^j(x);\;\;(j=\gamma,\gamma Z, Z),\\
 \left[F_{2p}^{\gamma}, F_{2p}^{\gamma Z}, F_{2p}^{Z} \right] &=& x \sum_{i=u,d,s}\left[e_i^2, 2 e_i g_V^i, \left(g_V^{i}\right)^2+\left(g_A^{i}\right)^2 \right]\left(q_i(x)+\bar{q}_i(x)\right),\nonumber\\
 \left[F_{3p}^{\gamma}, F_{3p}^{\gamma Z}, F_{3p}^{Z} \right]&=&\sum_{i=u,d,s}\left[0,2 e_i g_A^i, 2 g_V^i g_A^i \right]\left(q_i(x)-\bar{q}_i(x)\right),
\end{eqnarray}
It should be noted that in the QPM, the structure functions $F_{1p}^j(x)$ are given in terms of $F_{2p}^j(x)$ through Eq.~\ref{cg0} which is 
known as the Callan-Gross relation~\cite{Callan:1969uq}. 
The proton structure functions $F_{2p}^j(x)$ contain the sum of quark and antiquark densities, and therefore receive contributions 
from both valence and sea quarks. In contrast, in the case of parity violating proton structure functions $F_{3p}^{j}(x)$, 
$F_{3p}^{\gamma }(x)=0$, and $F_{3p}^{\gamma Z}(x)$ and $F_{3p}^{Z}(x)$ are proportional to the difference between
quark and antiquark densities, and thus receive contributions from the valence quarks only.

However, as we move away from the kinematic region of the Bjorken scaling to the region of low and moderate
values of $Q^2$ and $\nu$, the Callan-Gross relation~\cite{Callan:1969uq} is expected to be modified even at the leading order of pQCD. The higher order perturbative corrections
due to the quark-quark and quark-gluon
interactions are expected to give rise to the $Q^2$ dependent contribution to the structure functions. These corrections lead to the violation of 
Bjorken scaling, and the Callan-Gross relation is then expressed as~\cite{renton}:
\begin{equation}\label{cgmod}
 F_{1p}^{\gamma/\gamma Z}(x,Q^2)=\frac{1}{2x}\;\Big[(1+\gamma^2) \; F_{2p}^{\gamma/\gamma Z}(x,Q^2)-F_{Lp}^{\gamma/\gamma Z}(x,Q^2)\Big],\;\;\gamma^2=\frac{4 M^2 x^2}{Q^2},
\end{equation}
where $F_{Lp}(x,Q^2)$ is the longitudinal proton structure function that gives non-zero contributions at low and moderate
$Q^2$ ($F_{Lp}^{(LO)}(x,Q^2)\ne 0$), and is modified by the strong coupling constant $\alpha_s(Q^2)$ even at LO of pQCD~\cite{cooper_sarkar, Moch:2004xu}. 
We follow the prescription by Moch et al.~\cite{Moch:2004xu}, where the longitudinal structure function $F_{Lp}^{\gamma/\gamma Z}(x,Q^2)$ 
at LO and beyond it (Eqs.(1-11) in Ref.~\cite{Moch:2004xu}) has the contributions from gluons as well as quark contributions
through the coefficient functions, to evaluate $F_{1p}^{\gamma/\gamma Z}(x,Q^2)$ using Eq.~\ref{cgmod} for finite values of $Q^2$. In the evaluation of 
$F_{Lp}^{\gamma/\gamma Z}(x,Q^2)$ the gluonic and the quark terms associated with the coefficients functions get modified by the charge factors corresponding 
to the $\gamma$-exchange and $\gamma Z$ interference term. In literature, some authors~\cite{Hobbs:2008mm, Brady:2011uy} have also studied the
Callan-Gross relation~\cite{Callan:1969uq} in the presence of corrections at the
finite values of $Q^2$ and taking the non-zero values of $R_p^{\gamma/\gamma Z}(x,Q^2)$. For example, Hobbs et al.~\cite{Hobbs:2008mm}
studied the effect of non-zero value of $R_{p,d}^{\gamma/\gamma Z}(x,Q^2)$ on the electron spin asymmetry at LO, and Brady et al.~\cite{Brady:2011uy} 
have studied the dependence of $R_{p,d}^{\gamma/\gamma Z}(x,Q^2)$ and the asymmetry $A_{PV}^{(e)}(x,Q^2)$ on the different prescriptions 
of target mass corrections at NLO for the proton and deuteron targets.

Using Eq.~\ref{fnc} in Eq.~\ref{d2nc_comp}, the differential cross section for the polarized electron with 
helicity $\lambda_e$ can be written as:
\begin{small}
\begin{eqnarray}\label{dsig}
 \frac{d^2\sigma}{dx dy}&=& \big(\frac{2\pi y \alpha^2}{q^4}\big)\;\big(\frac{4 M E}{y}\big)\;\Big[ xy^2 \;F_{1p}^{\gamma }(x,Q^2)+(1-y)\;F_{2p}^{\gamma }(x,Q^2)\nonumber\\
 &&-\eta^{\gamma Z}\;(g_V^e-\lambda_e\;g_A^e)\;\Big\{ xy^2 \;F_{1p}^{\gamma Z}(x,Q^2)+(1-y)\;F_{2p}^{\gamma Z}(x,Q^2)+ \lambda_e x y \big(1-\frac{y}{2}\big) \;F_{3p}^{\gamma Z}(x,Q^2)\Big\} \nonumber\\
&& +(\eta^{\gamma Z})^2\;(g_V^e-\lambda_e\;g_A^e)^2\;\Big\{ xy^2 \;F_{1p}^{Z }(x,Q^2)+(1-y)\;F_{2p}^{Z }(x,Q^2)+\lambda_e  x y \big(1-\frac{y}{2}\big) \;F_{3p}^{Z }(x,Q^2)\Big\}\Big],
\end{eqnarray}
\end{small}
where $\lambda_e=+1(-1)$ correspond to the right handed (R) and left handed (L) polarized electron scattering from unpolarized proton.
For the numerical calculations, we use the MMHT PDFs parameterization~\cite{Harland-Lang:2014zoa} in the MSbar-scheme
and evolve the parton densities up to NNLO following Refs.~\cite{Moch:2004xu, Vermaseren:2005qc, Moch:2008fj, Zaidi:2019mfd}, where the proton structure functions
are defined as:
\begin{equation}\label{F_L}
x^{-1}F_{2,L,3}(x)=\sum_{f=q,g} C_{2,L,3}^{(n)}(x) \otimes f(x), 
\end{equation}
where $C_{2,L,3}$ are the coefficient functions for the quarks and gluons, the superscript $n=0,1,2,3....$ for
N$^{n}$LO, the symbol $\otimes$ represents the Mellin convolution and $f$ represents the quark and gluon distributions~\cite{Harland-Lang:2014zoa}.
Furthermore, we have incorporated TMC effect following Ref.~\cite{Schienbein:2007gr} and the twist-4 effect following Ref.~\cite{Dasgupta:1996hh, Stein:1998wr}.
At low $Q^2$, the proton structure functions are expressed in terms of powers of $1/Q^2$ (up to twist-4) in the operator product expansion as:  
\begin{equation}
F_{ip}(x,Q^2) = F_{ip}^{\tau = 2}(x,Q^2)
+ {H_{ip}^{\tau = 4}(x) \over Q^2}  \;\;\; i=1,2,
\label{eqn:ht}
\end{equation}
 where the first term ($\tau=2$) is known as the leading twist (LT) term, which incorporates the evolution 
of structure functions via perturbative QCD $\alpha_s(Q^2)$ corrections, and the second term $(\tau = 4)$ is  
the higher twist (twist-4) term which reflects the strength of multi-parton correlations (quarks and gluons inside a hadron are correlated). 

\subsection{Electron spin asymmetry $A_{PV}^{(e)}(x,Q^2)$}
The parity violating spin asymmetry for
the longitudinally polarized electron beam off unpolarized proton target is defined as~\cite{Hobbs:2008mm}:
\begin{equation}\label{asee}
  A_{PV}^{(e)}(x,Q^2)=\frac{\sigma_R(x,Q^2)-\sigma_L(x,Q^2)}{\sigma_R(x,Q^2)+\sigma_L(x,Q^2)},
\end{equation}
where the differential scattering cross sections for the scattering of right-handed ($\sigma_R(x,Q^2)$) and left-handed ($\sigma_L(x,Q^2)$) electron off an unpolarized 
proton target are given by:
\begin{equation}\label{xsec_spin}
 \sigma_{R(L)}(x,Q^2)= \frac{d^2\sigma}{dx dy}\Big(\lambda_e=+1(-1)\Big),
\end{equation}
where average is taken over the proton spin states $\lambda_p$.
\begin{equation}
 \frac{d^2\sigma^{0}}{dx dy}=\sigma_R(x,Q^2) + \sigma_L(x,Q^2)
\end{equation}

We evaluate the unpolarized differential scattering cross section ($\sigma_R(x,Q^2)+\sigma_L(x,Q^2)$) using Eq.~\ref{dsig} and 
summing over all possible spin orientations of the charged lepton beam and the target nucleon as well as 
the averaged differential cross sections ($\sigma_R(x,Q^2)-\sigma_L(x,Q^2)$) 
for the longitudinally polarized electron beam scattering off unpolarized nucleon target~\cite{Boughezal:2022pmb}, and by taking their ratio 
we obtain the electron asymmetry $A_{PV}^{(e)}(x,Q^2)$~\cite{Hobbs:2008mm}:
\begin{eqnarray}
\label{astt}
 A_{PV}^{(e)}(x,Q^2)&=&\frac{\sigma_R(x,Q^2)-\sigma_L(x,Q^2)}{\sigma_R(x,Q^2)+\sigma_L(x,Q^2)}=\frac{{\cal A}(x,Q^2)}{{\cal B}(x,Q^2)},\\
\textrm{where}~~ {\cal A}(x,Q^2)&=&|P_e|\Big[\eta^{\gamma Z}\Big\{ g_A^e\big(xy^2 \;F_{1p}^{\gamma Z}(x,Q^2)+(1-y)\;F_{2p}^{\gamma Z}(x,Q^2)\big)-g_V^e x y \big(1-\frac{y}{2}\big) \;F_{3p}^{\gamma Z}(x,Q^2)\nonumber\\
 &+& (\eta^{\gamma Z})^2\Big\{ -2 g_V^e g_A^e\big(xy^2 \;F_{1p}^{Z}(x,Q^2)+(1-y)\;F_{2p}^{ Z}(x,Q^2)\big)+\big((g_V^e)^2+(g_A^e)^2\big) x y \big(1-\frac{y}{2}\big) \;F_{3p}^{Z}(x,Q^2)\Big\}\Big],\nonumber\\
{\cal B}(x,Q^2)&=&  \Big\{ xy^2 \;F_{1p}^{ \gamma}(x,Q^2)+(1-y)\;F_{2p}^{\gamma}(x,Q^2)\Big\}\nonumber\\
 &&-\eta^{\gamma Z}\Big\{ g_V^e\big(xy^2 \;F_{1p}^{\gamma Z}(x,Q^2)+(1-y)\;F_{2p}^{\gamma Z}(x,Q^2)\big)-g_A^e x y \big(1-\frac{y}{2}\big) \;F_{3p}^{\gamma Z}(x,Q^2)\Big\}\nonumber\\
 &&+(\eta^{\gamma Z})^2\Big\{ \big((g_V^e)^2+(g_A^e)^2\big) \big(xy^2 \;F_{1p}^{Z }(x,Q^2)+(1-y)\;F_{2p}^{Z }(x,Q^2)\big)-2 g_V^e g_A^e x y \big(1-\frac{y}{2}\big) \;F_{3p}^{ Z}(x,Q^2)\Big\},\nonumber
\end{eqnarray}
where $|P_e|$ is the polarization vector for the longitudinally polarized electron beam.
It is important to note that in polarization experiments
like PVDIS experiments at JLab~\cite{Accardi:2023chb, Meziani:2024leh, JeffersonLabSoLID:2022iod}, EIC~\cite{AbdulKhalek:2021gbh, Boughezal:2022pmb} and 
EicC~\cite{Anderle:2021wcy}, the polarization of electron beam is about 80\%, i.e., $|P_e|=0.8$. Therefore, we have taken
$|P_e|=0.8$ for the numerical calculations of the electron asymmetry.

For $Q^2<< M_Z^2$, the contribution from the pure $Z-$exchange is suppressed by a factor of
$(\eta^{\gamma Z})^2$ (Eq.~\ref{coup}), which is of the order of $10^{-6}$ for $Q^2\simeq 10$ GeV$^2$, therefore, here on-wards, we will not
consider the contribution from the pure weak $Z-$channel.
The denominator, on the other hand, contains all the contributions, but is
dominated by the purely electromagnetic component.
Hence, the electron spin asymmetry (Eq.~\ref{astt}) can be written as~\cite{Hobbs:2008mm}:
\begin{eqnarray}\label{ase}
 A_{PV}^{(e)}(x,Q^2)&=&|P_e|\frac{\eta^{\gamma Z}\Big\{ g_A^e\big(xy^2 \;F_{1p}^{\gamma Z}(x,Q^2)+(1-y)\;F_{2p}^{\gamma Z}(x,Q^2)\big)-g_V^e x y \big(1-\frac{y}{2}\big) \;F_{3p}^{\gamma Z}(x,Q^2)\Big\}}{  \;\Big\{xy^2 \;F_{1p}^{ \gamma}(x,Q^2)+(1-y)\;F_{2p}^{ \gamma}(x,Q^2)\Big\}},
\end{eqnarray}
Alternatively, the above expression of the electron spin asymmetry given in Eq.~\ref{ase} may also be expressed in terms of the ratio of the 
longitudinal $F_{Lp}(x,Q^2)$ $\Big(=\Big[1+\frac{4 M^2 x^2}{Q^2}\Big] F_{2p}(x,Q^2)-2 x F_{1p}(x,Q^2)\Big)$
to the transverse $F_{Tp}(x,Q^2)$ ($=2 x F_{1p}(x,Q^2)$) structure functions as:
\begin{eqnarray}\label{aee0}
  A_{PV}^{(e)}(x,Q^2)&=&|P_e|\eta^{\gamma Z} \Big[ g_A^e \Big(\frac{F_{1p}^{\gamma Z}(x,Q^2)}{F_{1p}^{\gamma}(x,Q^2)} \Big)\;\frac{y^2+\frac{2}{r^2} (1-y)(1+R_p^{\gamma Z}(x,Q^2))}{y^2+\frac{2}{r^2} (1-y)(1+R_p^{\gamma}(x,Q^2))}\nonumber\\
  &-&  g_V^e \Big(\frac{F_{3p}^{\gamma Z}(x,Q^2)}{F_{1p}^{\gamma}(x,Q^2)}\Big)\;\frac{y\big(1-\frac{y}{2}\big)}{y^2+\frac{2}{r^2} (1-y)(1+R_p^{\gamma}(x,Q^2))}\Big],
\end{eqnarray}
where $R_p^{\gamma Z/\gamma}(x,Q^2)$ is given by
\begin{equation}\label{rlp}
 R_p^{\gamma Z/\gamma}(x,Q^2)=\frac{F_{Lp}^{\gamma Z/\gamma}(x,Q^2)}{2 x F_{1p}^{\gamma Z/\gamma}(x,Q^2)}=\frac{(1+\gamma^2) \; F_{2p}^{\gamma Z/\gamma}(x,Q^2)-2 x F_{1p}^{\gamma Z/\gamma}(x,Q^2)}{2 x F_{1p}^{\gamma Z/\gamma}(x,Q^2)},
\end{equation}
with $r^2=1+\gamma^2$.

The above expression (Eq.~\ref{aee0}) in a more compact form is given by~\cite{Hobbs:2008mm}:
\begin{eqnarray}\label{aee}
  A_{PV}^{(e)}(x,Q^2)&=&|P_e|\eta^{\gamma Z}\Big\{   Y_{1p}(x,Q^2) a_{1p}(x,Q^2) -   Y_{3p}(x,Q^2) a_{3p}(x,Q^2)\Big\},
\end{eqnarray}
where 
\begin{equation}\label{y13}
 Y_{1p}(x,Q^2)=\frac{y^2+2 (1-y)(1+R_p^{\gamma Z}(x,Q^2))/r^2}{y^2+2 (1-y)(1+R_p^{\gamma}(x,Q^2))/r^2};\;\;\;Y_{3p}(x,Q^2)=\frac{y\big(1-\frac{y}{2}\big)}{y^2+2 (1-y)(1+R_p^{\gamma}(x,Q^2))/r^2},
\end{equation}
 and 
 \begin{eqnarray}\label{a13}
 a_{1p}(x,Q^2)&=&g_A^e \Big(\frac{F_{1p}^{\gamma Z}(x,Q^2)}{F_{1p}^{\gamma}(x,Q^2)} \Big),\hspace{3 mm}
 a_{3p}(x,Q^2)=g_V^e \Big(\frac{F_{3p}^{\gamma Z}(x,Q^2)}{F_{1p}^{\gamma}(x,Q^2)} \Big).
\end{eqnarray}
In the LO of QCD, $a_{1p}(x)$ and $a_{3p}(x)$ are given by (using Eq.~\ref{a13})~\cite{Hobbs:2008mm}:
\begin{eqnarray}
 a_{1p}(x)&=&\frac{\sum_i\;C_{1i} e_i (q_i(x)+\bar q_i(x))}{\sum_i e_i^2 (q_i(x)+\bar q_i(x))},\hspace{3 mm}
 a_{3p}(x)=\frac{\sum_i\;C_{2i} e_i (q_i(x)-\bar q_i(x))}{\sum_i e_i^2 (q_i(x)+\bar q_i(x))},
 \label{a130}
\end{eqnarray} 
where $C_{1i}=2 g_A^e g_V^i$ and $C_{2i}=2 g_V^e g_A^i$ are the effective weak coupling constants~\cite{ParticleDataGroup:2024cfk}. In the 
standard model, the vector ($g_V^i$) and axial-vector ($g_A^i$) couplings of quarks are given by~\cite{Hobbs:2008mm} 
\begin{equation}
 g_V^u=-\frac{1}{2}+\frac{4}{3}\;\sin^2\theta_W\;;\;\;g_V^d=\frac{1}{2}-\frac{2}{3}\;\sin^2\theta_W\;;\;g_V^s=g_V^d\;;\;g_A^u=\frac{1}{2}\;;\;\;g_A^d=-\frac{1}{2}\;;\;g_A^s=g_A^d,
\end{equation}
leading to
\begin{equation}\label{c1u1}
 C_{1u}=\Big(\frac{1}{2}-\frac{4}{3}\;\sin^2\theta_W\Big)=0.19173, ~~~C_{1d}=\Big(-\frac{1}{2}+\frac{2}{3} \sin^2\theta_W\Big) =-0.34586
\end{equation}
and 
\begin{equation}\label{c2u2}
 C_{2u}=\Big(-\frac{1}{2}+2 \sin^2\theta_W \Big)=-0.0376, ~~~C_{2d}=\Big(\frac{1}{2}-2 \sin^2\theta_W \Big)=0.0376
\end{equation}
using $\sin^2\theta_W=0.2312$~\cite{ParticleDataGroup:2024cfk}.

It is important to note that in the standard model, the electron asymmetry $A_{PV}^{(e)}(x,Q^2)$ is dominated by the first term containing $a_{1p}(x,Q^2)$
as the second term containing $a_{3p}(x,Q^2)$ is suppressed by the 
vector coupling $g_V^e$ (-0.0376: Eq.~\ref{coups}) which is much smaller in magnitude than 
the axial-vector coupling $g_A^e$ (-0.5: Eq.~\ref{coups}). The values of effective weak coupling constants $C_{1u,1d}$ and $C_{2u,2d}$
have been obtained using Eqs.~\ref{c1u1}, \ref{c2u2}, and are used in Eq.~\ref{a130} to evaluate $a_{1p}(x,Q^2)$ and $a_{3p}(x,Q^2)$ .

In the Bjorken limit of $Q^2\to \infty$, where the Callan-Gross (CG) relation~\cite{Callan:1969uq} holds, the longitudinal structure function ($F_{Lp}^{\gamma Z/\gamma}(x,Q^2)$)
becomes zero (i.e., $F_{Lp}^{\gamma Z/\gamma}(x,Q^2)\to 0$), and 
\begin{equation}
 \gamma \to 0, ~r\to 1~\textrm{and} ~R_p^{\gamma/\gamma Z} \to 0\;\;\textrm{(following Eqs.~\ref{cgmod} and \ref{rlp})}.
\end{equation}
In this case (using Eq.~\ref{y13}), $Y_{1p}$ approaches unity, i.e., $Y_{1p} \to 1$,
$Y_{3p} \to \frac{1}{2}\;\frac{y(2-y)}{y^2+2(1-y)}\equiv f(y)$, and the asymmetry $A_{PV}^{(e)}(x,Q^2)$ using Eqs.~\ref{aee0}-\ref{a13} reduces to:
\begin{equation}\label{aes0}
   A_{PV}^{(e)}(x)=|P_e|\eta^{\gamma Z}\Big\{g_A^e \frac{F_{2p}^{\gamma Z}(x)}{F_{2p}^{\gamma}(x)} - f(y)\; g_V^e \frac{2 x F_{3p}^{\gamma Z}(x)}{F_{2p}^{\gamma}(x)}   \Big\}
\end{equation}
For low and moderate values of $Q^2$, $F_{Lp}(x,Q^2)$ gives non-zero contribution, hence, it becomes important to know $R_p^\gamma(x,Q^2)$ 
and $R_p^{\gamma Z}(x,Q^2)$, and test whether $R_p^{\gamma}$ is different
from $R_p^{\gamma Z}$ or not? Moreover, the study of the effects of higher order perturbative and nonperturbative corrections 
on $R_p^{\gamma Z/\gamma}(x,Q^2)$ as well as the applicability of 
the Callan-Gross relation would be helpful in improving our understanding of the parity violating electron asymmetry and extraction of the 
weak mixing angle $\theta_W$.

\section{Results and discussion}\label{results}
\begin{figure}
 \includegraphics[height=8 cm, width=\textwidth]{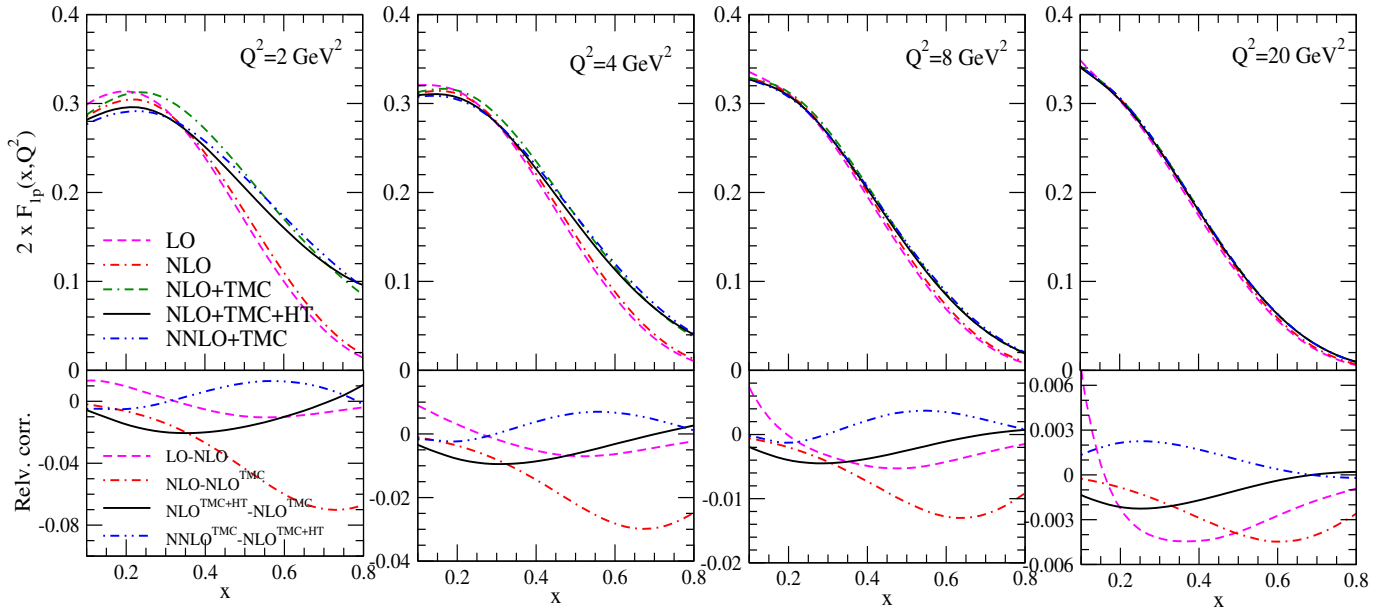}
 \caption{Results for the unpolarized proton electroweak structure functions $2 x F_{1p} (x,Q^2)$ (top panel) vs $x$ at the different values of $Q^2$ 
 incorporating the contributions from $\gamma$ and $\gamma Z$ exchanges.
 The evaluation is performed at LO, NLO and NNLO without and with the nonperturbative effects like TMC and HT (as mentioned in the legends of the figure).
 These results are obtained using MMHT14 PDFs parameterization~\cite{Harland-Lang:2014zoa} in the MSbar scheme. In the bottom panel of the figure relative 
 corrections are presented to explicitly show the contributions from perturbative and nonperturbative effects (as mentioned in the legends of the figure).}
 \label{res1a}
\end{figure}

\begin{figure}
 \includegraphics[height=8 cm, width=\textwidth]{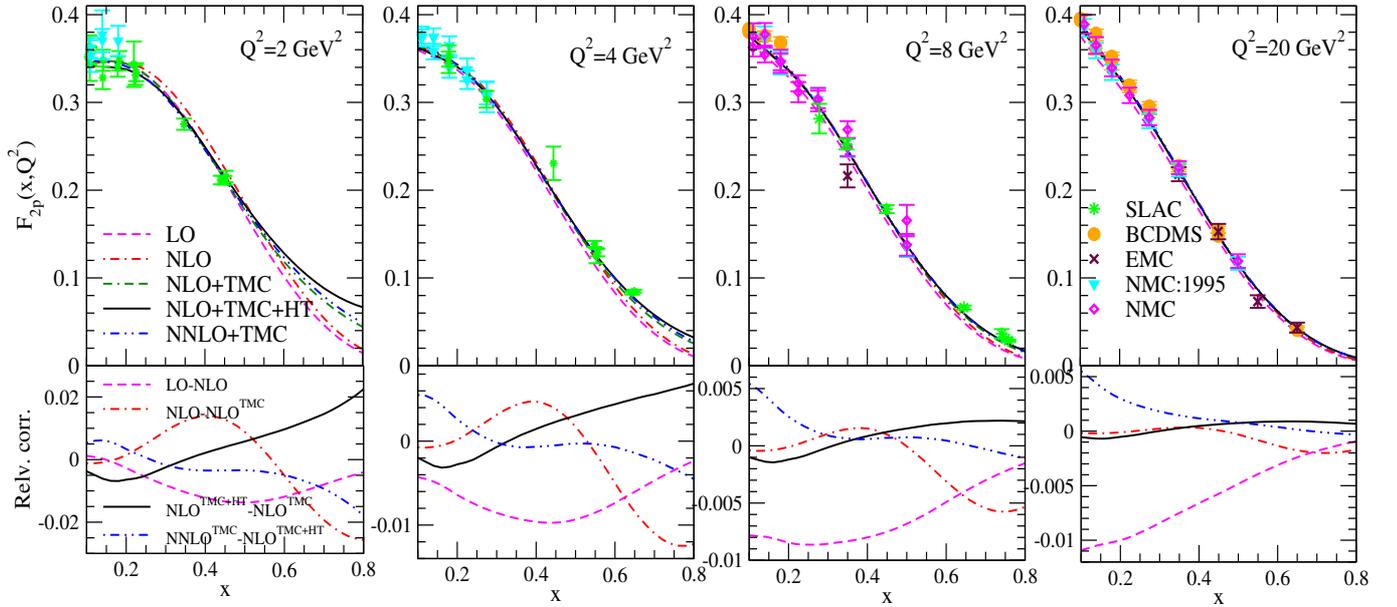}
 \caption{Results for the unpolarized proton electroweak structure functions $F_{2p} (x,Q^2)$ (top panel) vs $x$ at the different values of $Q^2$ 
 incorporating the contributions from $\gamma$ and $\gamma Z$ exchanges. These results are compared with the experimental data corresponding to the interaction
 via pure photon exchange~\cite{Whitlow:1991uw, Benvenuti:1989rh, Arneodo:1996rv, Aubert:1985fx}. 
 The lines have the same meaning as in Fig.~\ref{res1a}. }
 \label{res1b}
\end{figure}

\begin{figure}
 \includegraphics[height=8 cm, width=\textwidth]{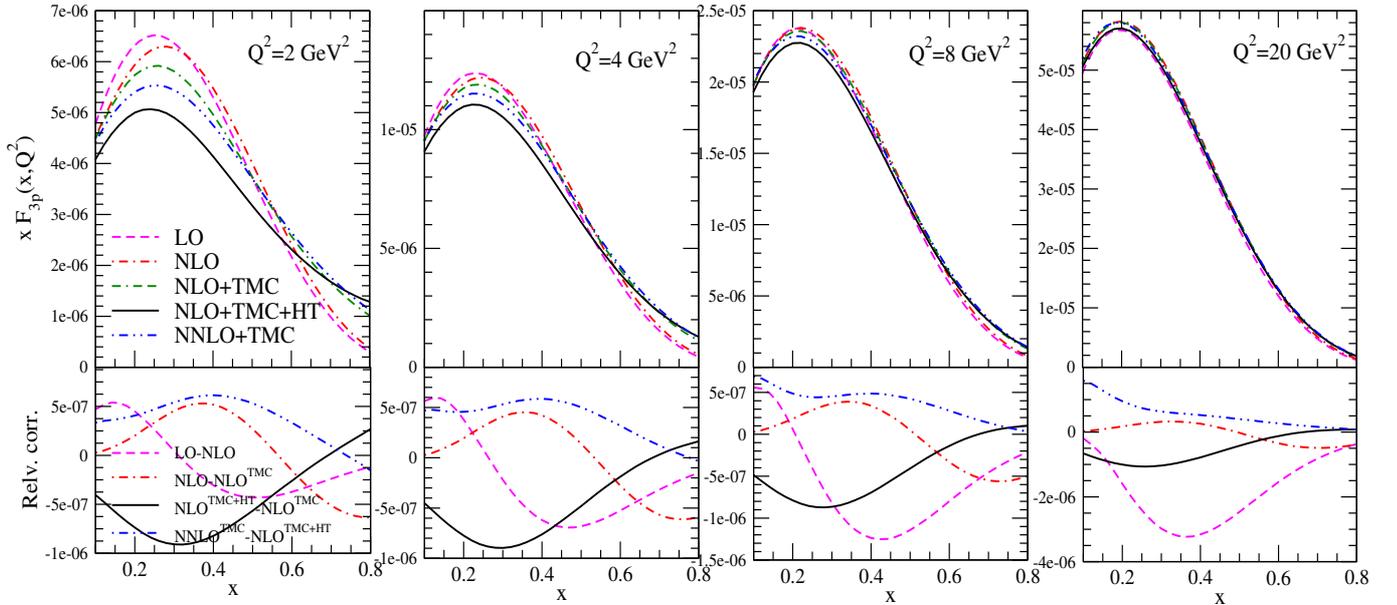}
 \caption{Results for the unpolarized proton electroweak structure functions $x F_{3p} (x,Q^2)$ (top panel) vs $x$ at the different values of $Q^2$ 
 incorporating the contributions from $\gamma$ and $\gamma Z$ exchanges ($x F_{3p}^{\gamma} (x,Q^2)=0$). 
 The lines have the same meaning as in Fig.~\ref{res1a}.  }
 \label{res1c}
\end{figure}
%
We present the numerical results for the electroweak proton structure functions $F_{ip}(x,Q^2);~i=1-3$ defined in Eq.~\ref{fnc} 
which include the contributions from 
$\gamma-$exchange and $\gamma Z$ interference terms and neglect the contributions from $Z$ exchange which are quite 
small due to $(\eta^{\gamma Z})^2$. The results are also
presented for the ratios $r^{\gamma Z/\gamma}(x,Q^2)=\frac{F_{2p}^{\gamma Z/\gamma}(x,Q^2)}{2 x F_{1p}^{\gamma Z/\gamma}(x,Q^2)}$,
$R^{\gamma Z/\gamma}(x,Q^2)=\frac{F_{Lp}^{\gamma Z/\gamma}(x,Q^2)}{2 x F_{1p}^{\gamma Z/\gamma}(x,Q^2)}$, 
$a_{1p}(x,Q^2)$, the coefficients $Y_{1p}(x,Q^2)$ and $Y_{3p}(x,Q^2)$, and discuss their contributions
in the calculations of the electron beam spin asymmetry $A_{PV}^{(e)}(x,Q^2)$ (defined through Eq.~\ref{y13}). We also discuss the impact of
perturbative and nonperturbative QCD corrections on various structure functions of the nucleon and the PV electron asymmetry $A_{PV}^{(e)}(x,Q^2)$ for
which the numerical results are presented at the beam energies relevant to
the experiments at JLab and its upgrade as well as the proposed experiments at EIC and EicC. The numerical calculations are performed by evolving 
the parton densities (PDFs) up to the 
next-to-next-to-leading order (NNLO) in the 3-flavor ($u,d,s$) MSbar scheme incorporating the 
nonperturbative target mass correction (TMC) and the higher twist effect (HT: twist-4). All the numerical calculations are performed 
in the kinematic region of $0.1 \le x \le 0.8$ and $Q^2\ge 1$ GeV$^2$.

\subsection{Structure functions}
In the top panel of Figs.~\ref{res1a}-\ref{res1c}, we have respectively shown the results for the electroweak structure functions, i.e., $2x F_{1p} (x,Q^2)$,
$F_{2p} (x,Q^2)$ and $x F_{3p} (x,Q^2)$, calculated using Eq.~\ref{fnc}.
These results are shown for the specific values of $Q^2$ viz. $Q^2=2,4,8$ and 20 GeV$^2$ (left to right panel). 
The proton structure functions $2x F_{1p} (x,Q^2)$ and $F_{2p} (x,Q^2)$
are dominated by the contribution from the photon($\gamma$)-exchange $2x F_{1p}^{\gamma} (x,Q^2)$ and $F_{2p}^{\gamma} (x,Q^2)$ as
the contribution of $2x F_{1p}^{\gamma Z}(x,Q^2)$ and 
$F_{2p}^{\gamma Z}(x,Q^2)$ are very small (suppressed by a factor of $10^{-3}$). In the case of the parity violating structure function
$x F_{3p} (x,Q^2)$, the contribution arises solely from the $\gamma Z$ interference, i.e., $x F_{3p}^{\gamma Z}(x,Q^2)$. In the bottom panel 
of these Figs.~\ref{res1a}-\ref{res1c}, we have also explicitly shown the contributions from the higher order perturbative and nonperturbative effects
by plotting the difference in the results when the different corrections are incorporated.
We find that:
\begin{itemize}
 \item The corrections due to the inclusion of NLO terms in $2x F_{1p} (x,Q^2)$ is small in the region of low $x$ and low $Q^2$ and becomes
further smaller with the increase in $x$ and $Q^2$. For example, at $Q^2=2$ GeV$^2$, the reduction due to the inclusion of NLO terms 
is about $3\%$ at $x=0.2$ and an enhancement of $2\%$ at $x=0.5$ relative to the results obtained at LO, however, this difference becomes 
very small at high $x$. Furthermore, the difference decreases with the increase in $Q^2$.
\item In the case of $F_{2p} (x,Q^2)$ the corrections due to the inclusion of next-to-leading order terms is comparatively smaller in the region 
of low $x$, i.e., about $<1\%$ at $x=0.2$ than what has been observed for $2x F_{1p} (x,Q^2)$,
and about $8\%$ at $x=0.5$ for $Q^2=2$ GeV$^2$. However, the dependence of $F_{2p} (x,Q^2)$ on $x$ and $Q^2$ is qualitatively the same as that of $2x F_{1p} (x,Q^2)$.
\item  In $x F_{3p} (x,Q^2)$, when NLO terms are incorporated there is a significant effect, and the results change considerably from LO. 
This effect is $x$ dependent. For example, there is
a reduction of $7\%$ at $x=0.2$ and an enhancement of $12\%$ at $x=0.5$ from the results obtained at LO. 
\item The TMC effect (double dash-dotted line) enhances $2x F_{1p} (x,Q^2)$ from the results obtained without including it at NLO (dash-dotted line).
For example, the enhancement due to the incorporation of TMC effect is $26\%$ at $x=0.5$, $35\%$ at $x=0.6$
and $78\%$ at $x=0.7$ for $Q^2=2$ GeV$^2$, and for $Q^2=4$ GeV$^2$ this enhancement is about 
$6\%$ at $x=0.5$, $13\%$ $x=0.6$ and $\sim 50\%$ at $x=0.7$. On the other hand, the enhancement due to the inclusion of 
TMC effect is comparatively smaller in the case of $F_{2p} (x,Q^2)$ which is about 
$5\%$ at $x=0.5$, $4\%$ at $x=0.6$ and $26\%$ at $x=0.7$ for $Q^2=2$ GeV$^2$, and becomes
about $1\%$ at $x=0.5$, $5\%$ $x=0.6$ and $21\%$ at $x=0.7$ for $Q^2=4$ GeV$^2$. With the further increase in $Q^2$, say $Q^2=8$ GeV$^2$,
the difference between these two results (with and without TMC effect) is very small for both $2x F_{1p} (x,Q^2)$ and $F_{2p} (x,Q^2)$.

\item In the case of $x F_{3p} (x,Q^2)$ the inclusion of TMC effect leads to a reduction of about 
$8\%~(4\%)$ at $x=0.3$ and an enhancement of about $30\%~(25\%)$ at $x=0.7$ for $Q^2=2 (4)$ GeV$^2$. 
We observe that the TMC effect decreases with increasing $Q^2$ in all the three proton structure functions namely 
$2x F_{1p} (x,Q^2)$, $F_{2p} (x,Q^2)$ and $x F_{3p} (x,Q^2)$.
\item The inclusion of the higher twist effect (solid line) in $xF_{3p} (x,Q^2)$, leads to a further change in the numerical
results relative to the case with TMC effect only (double dash-dotted line) at NLO. These changes depend on both $x$ and $Q^2$. For example,
in $2 x F_{1p} (x,Q^2)$ for $Q^2=2$ (4) GeV$^2$ there is a reduction 
of about 8\% (3-4\%) in the region $0.3\le x\le 0.5$ and an enhancement of about 4\% (2\%) at $x=0.75$ when the 
HT effect is included along with the TMC effect. We observe that, for $Q^2=2$ (4) GeV$^2$, the enhancement 
in $F_{2p} (x,Q^2)$ due to the inclusion of the HT effect is almost negligible for $x<0.5$, while it
is about 3\% (2\%) at $x=0.5$, and increases to 28\% (15\%) at $x=0.75$. In contrast, in the case of $F_{3p} (x,Q^2)$, the higher twist effect 
leads to a reduction of about 16\% (8\%) at $x=0.3$ and 15\% (7\%) at $x=0.5$ along with an enhancement of about 9\% (5\%) at $x=0.75$. 
The higher twist correction in the region of low to intermediate $x$ is observed to be more pronounced in $F_{3p} (x,Q^2)$
than in $2 x F_{1p} (x,Q^2)$ and $F_{2p} (x,Q^2)$. 
\item We have also compared our numerical results at NLO including TMC and HT corrections, with the results at NNLO when only the TMC
effect is included (dash-double dotted line). It may be observed that the results for $2x F_{1p} (x,Q^2)$ 
and $F_{2p} (x,Q^2)$ at NNLO with TMC effect are similar to the NLO results with TMC and HT corrections. 
\item The numerical results for $x F_{3p} (x,Q^2)$ at NLO incorporating both the TMC and HT corrections have significant differences from the results 
evaluated at NNLO with TMC effect only at the lower values of $Q^2$, where the nonperturbative QCD corrections become significant. Quantitatively, 
the difference between these two results is about $8\%$ at $x=0.2$, $11\%$ at $x=0.3$ and $16\%$ at $x=0.6$ for $Q^2=2$ GeV$^2$, however, 
for $Q^2=8$ GeV$^2$ it becomes $4\%$ at $x=0.2$, $5\%$ at $x=0.3$ and $8\%$ at $x=0.6$. The difference between the results 
at NLO with TMC and HT effects and the results at NNLO with TMC effect only becomes quite small with increasing $Q^2$.
\end{itemize}

We have compared our numerical results for $F_{2p} (x,Q^2)$ with the available experimental data. Our results show reasonable agreement 
with measurements from the EMC, NMC, BCDMS and SLAC experiments~\cite{Whitlow:1991uw, Benvenuti:1989rh, Arneodo:1996rv, Aubert:1985fx}. 

\subsection{Callan Gross (CG) relation}
\begin{figure}
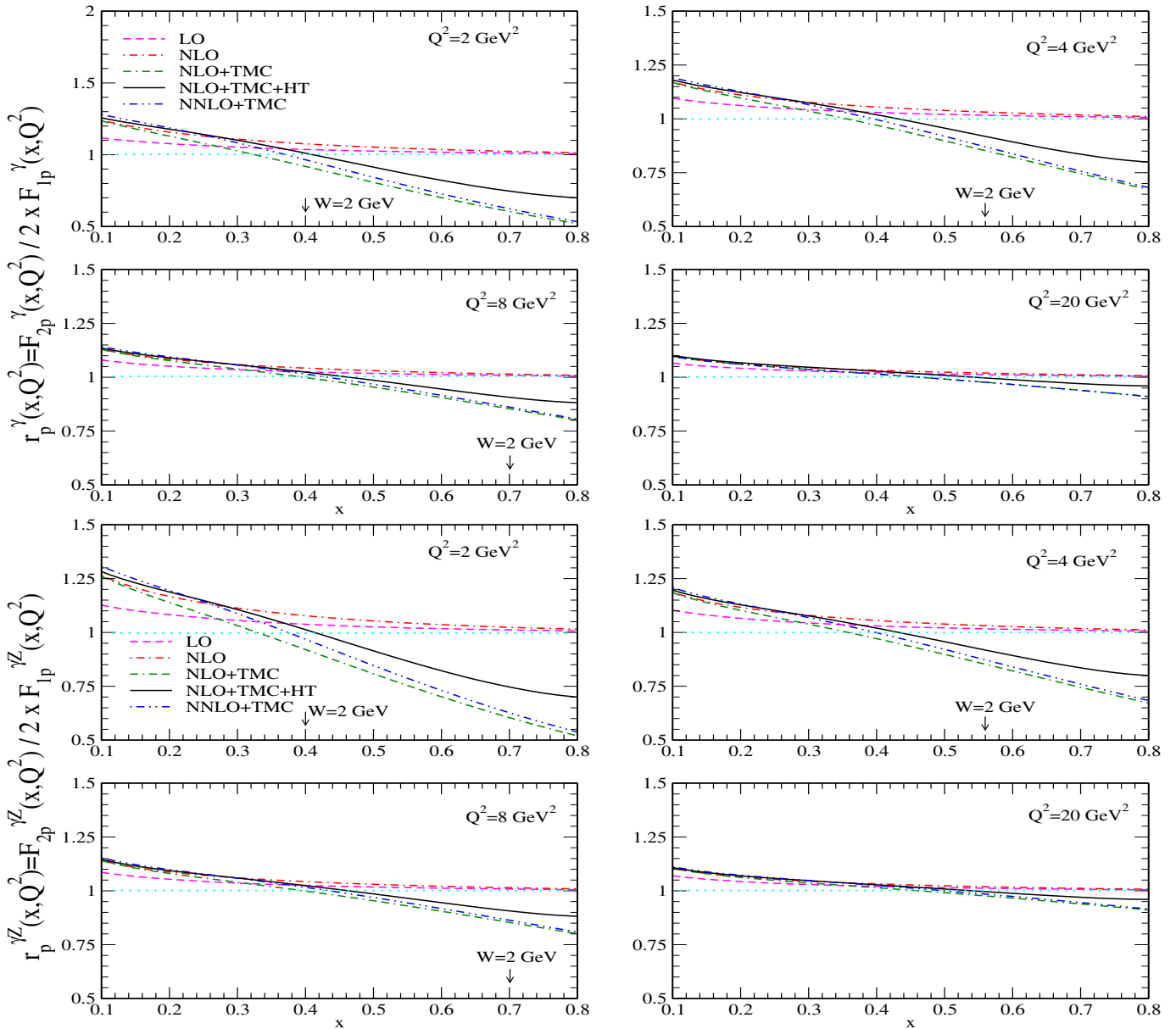

 \includegraphics[height=8 cm, width=\textwidth]{f2over2xf1_gamma_ratio.eps}
 \includegraphics[height=8 cm, width=\textwidth]{f2over2xf1_gammaz_ratio.eps}
 \caption{Results for the ratio $r_{p}^{\gamma }(x,Q^2)=\frac{F_{2p}^{\gamma}(x,Q^2)}{2 x F_{1p}^{\gamma}(x,Q^2)}$ (top panel) 
 and $r_{p}^{\gamma Z}(x,Q^2)=\frac{F_{2p}^{\gamma Z}(x,Q^2)}{2 x F_{1p}^{\gamma Z}(x,Q^2)}$ (bottom panel) vs $x$ at the different values of 
 $Q^2$. The evaluation is performed at LO, NLO and NNLO without and with the nonperturbative effects like TMC and HT (as mentioned in the legends of the figure). These 
 results are obtained using MMHT14 PDFs parameterization~\cite{Harland-Lang:2014zoa} in the MSbar scheme. }
 \label{res2}
\end{figure}

In Fig.~\ref{res2}, we have shown the results for the ratio $r_{p}^{\gamma }(x,Q^2)=\frac{F_{2p}^{\gamma}(x,Q^2)}{2 x F_{1p}^{\gamma}(x,Q^2)}$ (top panel) and 
$r_{p}^{\gamma Z}(x,Q^2)=\frac{F_{2p}^{\gamma Z}(x,Q^2)}{2 x F_{1p}^{\gamma Z}(x,Q^2)}$ (bottom panel) as a function of $x$ at the different values of $Q^2$ viz., $Q^2=2$ GeV$^2$, 4 GeV$^2$,
8 GeV$^2$ and 20 GeV$^2$. These results are relevant to test the validity of the Callan-Gross relation~\cite{Callan:1969uq} at 
low $Q^2$ in the leading order of QCD, 
as well as beyond the leading order at both lower and higher values of $Q^2$, with or without the inclusion of various corrections due to the nonperturbative 
and higher order perturbative effects. 
This analysis has been performed for both the electromagnetic sector proton structure 
functions and the weak-electromagnetic interference sector proton structure
functions.

It may be observed that:
\begin{itemize}
 \item The Callan-Gross relation does not hold good even at the leading order of perturbative QCD, in the kinematic region of low to mid $x$
at all values of $Q^2$ considered here. We find that at LO, the ratio $r_{p}^{\gamma}(x,Q^2)$ approaches CG limit, i.e., 
unity ($F_{2p}^{\gamma}(x)=2x F_{1p}^{\gamma}(x);~F_{Lp}^{\gamma}(x)=0$) for all values of $Q^2$ if $x \gtrsim 0.5$, 
implying that for $x \gtrsim 0.5$ Callan-Gross relation is valid. However, at low $x<0.5$ the ratio $r_{p}^{\gamma }(x,Q^2)$ significantly deviates from CG limit.
For example, at $Q^2=2$ GeV$^2$, we observe a deviation of about $10\%$ at $x=0.1$ and $2\%$ at $x=0.5$
from CG limit, and the deviation becomes smaller with the increase in $Q^2$. This is due to gluonic contributions, which
result in a finite value of the longitudinal structure function $F_{Lp}^{\gamma}(x,Q^2)$~\cite{Moch:2004xu}, contributing to the evaluation of
$F_{1p}^{\gamma}(x,Q^2)$ using Eq.~\ref{cgmod}. 
\item With the inclusion of NLO terms, the deviation from CG limit becomes more significant in the region of low to intermediate $x$. 
Quantitatively, the enhancement in the results for $r_{p}^{\gamma }(x,Q^2)$ evaluated at NLO compared to the results obtained at LO is about 
$11\%$ at $x=0.1$, $7\%$ at $x=0.2$ and $5\%$ at $x=0.3$ for $Q^2=2$ GeV$^2$ which becomes $7\%$ at $x=0.1$, $4\%$ at $x=0.1$ 
and $3\%$ at $x=0.3$ for $Q^2=4$ GeV$^2$. 
\item The inclusion of the TMC effect in $r_{p}^{\gamma }(x,Q^2)$ leads to an enhancement in the region of low $x (<0.4)$, while it causes 
a significant reduction from CG limit
in the region of intermediate to high $x$ ($0.4 \le x \le 0.8$). Overall, it results in a reduction relative to the case
without the inclusion of the TMC effect at NLO. For example, at $Q^2=2$ GeV$^2$ the reduction in the ratio $r_{p}^{\gamma }(x,Q^2)$ as compared to the results obtained without TMC effect 
at NLO is about $3\%$ at $x=0.2$, $15\%$ at $x=0.4$, $33\%$ at $x=0.6$ and about $50\%$ at $x=0.8$
which becomes $1\%$ at $x=0.2$, $8\%$ at $x=0.4$, $20\%$ at $x=0.6$ and about $34\%$ at $x=0.8$ for $Q^2=4$ GeV$^2$. Hence, the inclusion of TMC effect,
increases $r_{p}^{\gamma }(x,Q^2)$ with increasing $x$, and decreases with increasing $Q^2$.
\item The inclusion of higher twist corrections along with the TMC effect at NLO leads to an enhancement in the ratio $r_{p}^{\gamma }(x,Q^2)$,
relative to the results with TMC effect only, across the entire kinematic range of $x$ and $Q^2$. Quantitatively, 
this enhancement is about $10\%$ ($5\%$) at $x=0.4$, $15\%$ ($8\%$) at $x=0.6$ and $36\%$ ($20\%$) 
at $x=0.8$ for $Q^2=2$ (4) GeV$^2$. To conclude, both the TMC effect and the HT effect play an important role in the region of high $x$ and low $Q^2$,
and becomes small at low $x$ and high values of $Q^2$.
\item When the results are obtained at NNLO it has been observed that the difference between
the results obtained at NNLO and NLO is small in the entire region of Bjorken $x$. For example, the enhancement in the
ratio $r_{p}^{\gamma }(x,Q^2)$ is about 3-5\% for $x\le 0.4$ and $Q^2\le 4$ GeV$^2$ when NNLO terms are taken into account, and it further reduces with the increase in $x$ and $Q^2$.

Furthermore, the results at NLO with TMC and HT corrections are similar to the results at NNLO with TMC effect only in the region of low to intermediate $x$. However, at high $x$, 
there is a significant difference between them, especially at low values of $Q^2$. 
\end{itemize}

In the case of $\gamma Z$-interference, the deviation in the ratio $r_{p}^{\gamma Z}(x,Q^2)$ from CG limit is similar
to the case of $r_{p}^{\gamma}(x,Q^2)$. For example, for $Q^2=2$ GeV$^2$ the ratio obtained at the leading order 
deviates about $12\%$ at $x=0.1$ and $\sim 3\%$ at $x=0.5$ from CG limit while at NLO this deviation increases and becomes $25\%$ and $5\%$, respectively 
for $x=0.1$ and $x=0.5$. This deviation becomes almost negligible with the increasing $x(>0.7)$. 
It is also important to mention that, in the present case the contributions from
the TMC and HT corrections are similar to the case of $r_{p}^{\gamma}(x,Q^2)$. Quantitatively, the reduction in the ratio due to the
inclusion of TMC effect at NLO is about $14\% (8\%)$ at $x=0.5$ and $27\% (17\%)$ at $x=0.7$ 
when $Q^2=4$ (8) GeV$^2$, compared to the results evaluated at NLO without the TMC effect. Furthermore, the enhancement in the ratio obtained
by incorporating the HT effect together with the target mass corrections, relative to the results with the TMC effect only, is 
about 7\% (3\%) and 12\% (7\%) at $x=0.5$ and $x=0.7$ for $Q^2=4$ (8) GeV$^2$, respectively. We also find that the contribution
from NNLO terms to the ratio $r_{p}^{\gamma Z}(x,Q^2)$ is about $2\%-4\%$ for $0.1\le x\le 0.8$,
which is similar to the case of photon exchange.

This shows that the violation of the Callan-Gross relation~\cite{Callan:1969uq} may be due to the corrections at the finite values of $Q^2$ arising from
the scaling violation, and the non-zero value of longitudinal structure function $F_{Lp}^{\gamma /\gamma Z}(x,Q^2)$ even at the leading order 
of pQCD contributing to $2 x F_{1p}^{\gamma /\gamma Z}(x,Q^2)$. The inclusion of 
higher order perturbative corrections up to NNLO as well as the nonperturbative corrections like TMC and HT which are
found to be important in the present kinematic region of $x$ and $Q^2$ for both the photon exchange and $\gamma Z$ exchange terms. 

\begin{figure}
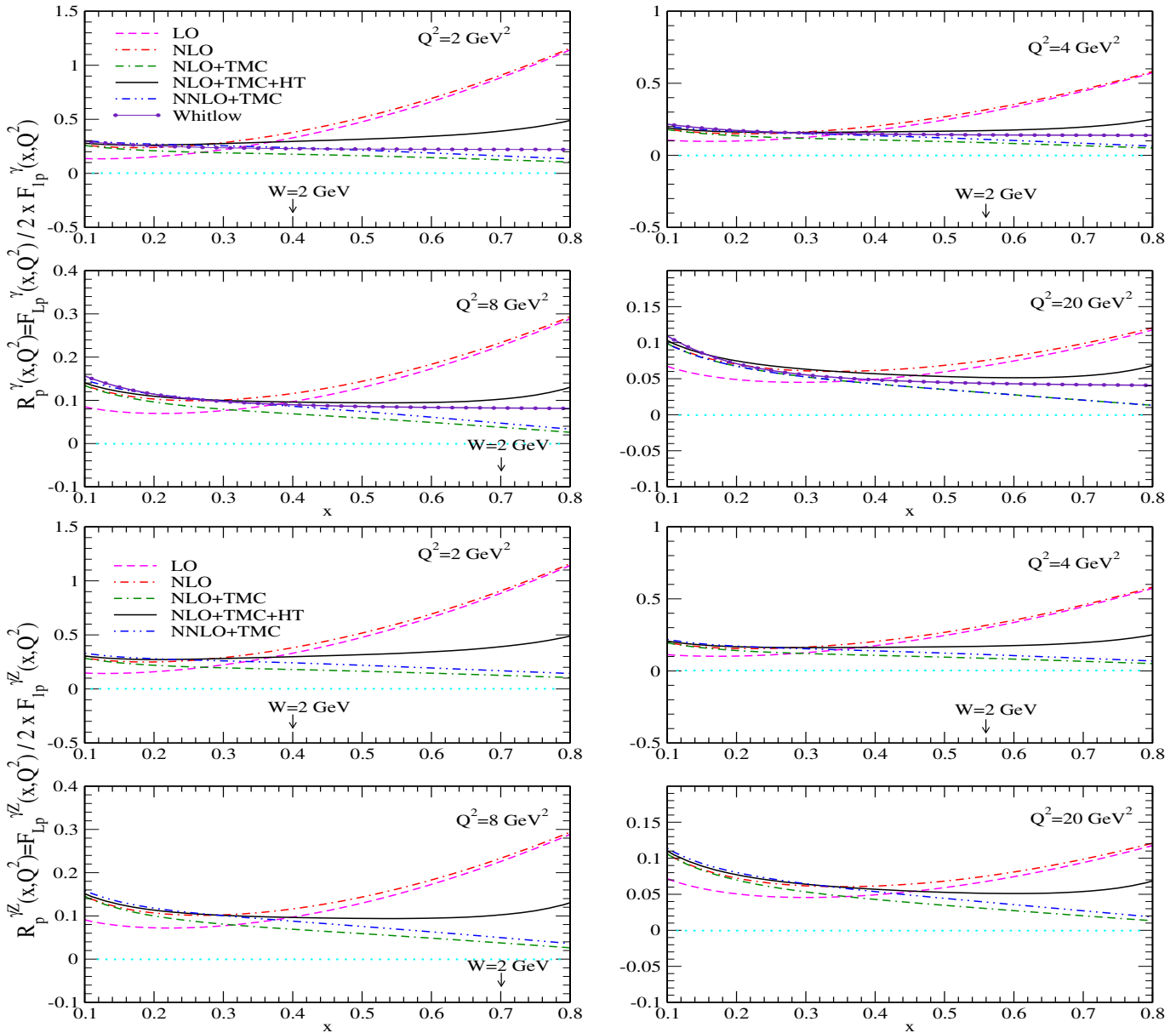

 \includegraphics[height=8 cm, width=\textwidth]{flover2xf1_gamma_ratio.eps}
 \includegraphics[height=8 cm, width=\textwidth]{flover2xf1_gammaz_ratio.eps}
 \caption{Results for the ratio $R_p^{\gamma}(x,Q^2)=\frac{F_{Lp}^{\gamma}(x,Q^2)}{2 x F_{1p}^{\gamma}(x,Q^2)}$ (top panel) 
 and $R_p^{\gamma Z}(x,Q^2)= \frac{F_{Lp}^{\gamma Z}(x,Q^2)}{2 x F_{1p}^{\gamma Z}(x,Q^2)}$ (bottom panel) vs $x$ at different values of 
 $Q^2$. The evaluation is performed at LO, NLO and NNLO without and with the nonperturbative effects like TMC and HT (as mentioned in the legends of the figure). These 
 results are obtained using MMHT14 PDFs parameterization~\cite{Harland-Lang:2014zoa} in the MSbar scheme. The results of $R_p^{\gamma}(x,Q^2)$ are also
 compared with the results obtained using the parameterization given by Whitlow et al.~\cite{Whitlow:1990gk, Whitlow:1991uw}.}
 \label{res3}
\end{figure}

\begin{figure}
 \includegraphics[height=8 cm, width=\textwidth]{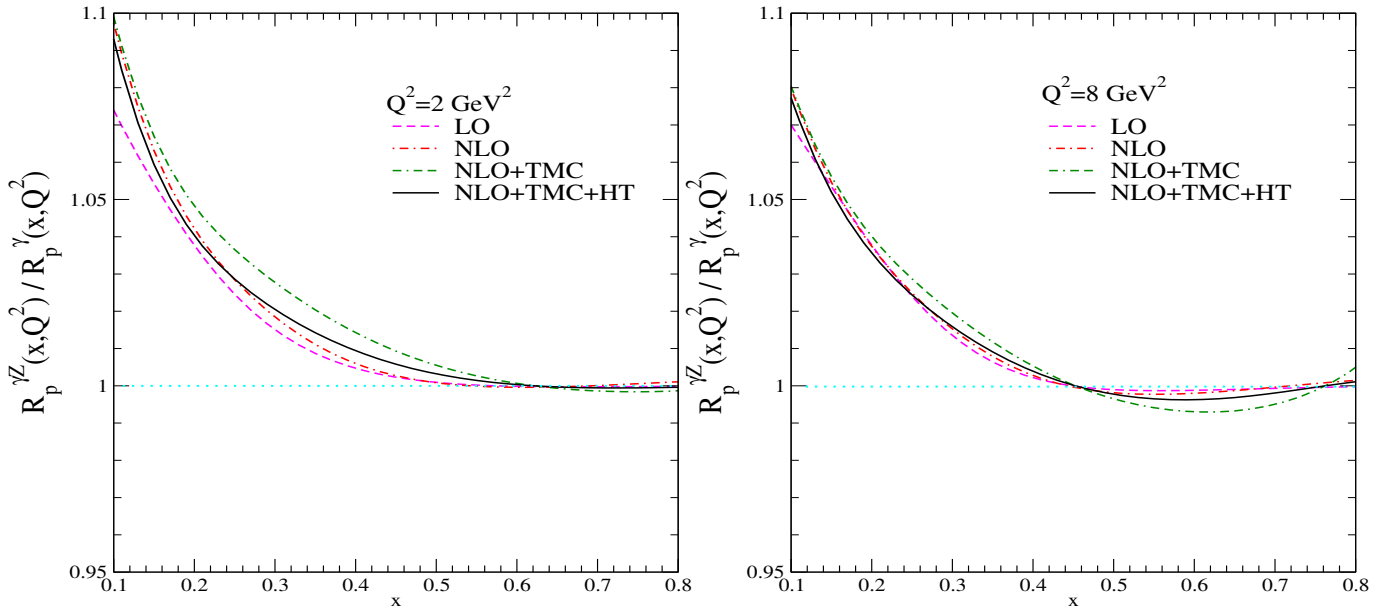}
 \caption{Results for the ratio $\frac{R_p^{\gamma Z}(x,Q^2)}{R_{p}^{\gamma}(x,Q^2)}$ vs $x$ at $Q^2=2$ GeV$^2$ (left panel)
 and $Q^2=8$ GeV$^2$ (right panel). The evaluation is performed up to NLO without and with the nonperturbative effects like TMC and HT
 (as mentioned in the legends of the figure). These 
 results are obtained using MMHT14 PDFs parameterization~\cite{Harland-Lang:2014zoa} in the MSbar scheme.}
 \label{res4}
\end{figure}

The main source of the violation of Callan-Gross relation~\cite{Callan:1969uq} in $r_{p}^{\gamma, \gamma Z}(x,Q^2)$ is due to the non-zero contribution from the longitudinal structure function 
$F_{Lp}^{\gamma/\gamma Z}(x,Q^2)$ which vanishes in the Callan-Gross limit. 
In our earlier work~\cite{Zaidi:2019mfd}, we have reported the 
numerical results for $R_{p}^{\gamma}(x,Q^2)=\frac{F_{Lp}^{\gamma}(x,Q^2)}{2 x F_{1p}^{\gamma}(x,Q^2)}$ calculated in this model 
in the case of unpolarized electron scattering from the unpolarized proton target via photon exchange, at the fixed
values of $x$ as a function of $Q^2$, incorporating the perturbative corrections up to NNLO along with the TMC and HT effects, and found these results to be
in reasonable agreement with the experimental data from SLAC. In literature, some phenomenological groups have also studied the deviation
of $R_{p}^{\gamma}(x,Q^2)$ from its Bjorken limit by studying the $Q^2$ dependence of $F_{Lp}^\gamma(x,Q^2)$ 
in the kinematic region of finite $Q^2$~\cite{Whitlow:1991uw, Whitlow:1990gk, Bodek:2010km, Christy:2007ve, Bosted:2015qc, 
 Abe:1998ym, Dasu:1988ms}.
 Among them the most widely used parameterization for the ratio $R_{p}^{\gamma}(x,Q^2)$
 is given by Whitlow et al.~\cite{Whitlow:1991uw, Whitlow:1990gk}. Furthermore, the ratios $R_{p}^{\gamma}(x,Q^2)=\frac{F_{Lp}^{\gamma}(x,Q^2)}{2 x F_{1p}^{\gamma}(x,Q^2)}$ and 
$R_{p}^{\gamma Z}(x,Q^2)=\frac{F_{Lp}^{\gamma Z}(x,Q^2)}{2 x F_{1p}^{\gamma Z}(x,Q^2)}$ for $\gamma-$exchange and $\gamma Z$ interference
terms, respectively, have also been discussed in some detail earlier by Hobbs et al.~\cite{Hobbs:2008mm} at LO, and Brady et al.~\cite{Brady:2011uy} at NLO.
In Fig.~\ref{res3}, we have presented the ratio $R_{p}^{\gamma}(x,Q^2)$ and $R_{p}^{\gamma Z}(x,Q^2)$ without and with the effect of the higher 
order perturbative corrections up to NNLO and including nonperturbative QCD corrections like TMC and HT (twist-4)
in a wide range of $x$ ($0.1\le x \le 0.8$) and $Q^2$ ($2 \le Q^2 \le 20$ GeV$^2$). Moreover, in Fig.~\ref{res3}, we have compared our numerical results 
for $R_{p}^{\gamma}(x,Q^2)$ with the results obtained by using the phenomenological parameterization by Whitlow et al.~\cite{Whitlow:1991uw, Whitlow:1990gk}.

At the leading order of pQCD, $R_{p}^{\gamma}(x,Q^2)$ is non-zero, 
due to gluonic corrections at low and moderate values of $Q^2$ (dashed line), as shown in Fig.~\ref{res3}.
From the figure, it may be noticed that $R_{p}^{\gamma}(x,Q^2)$ increases with the increase in $x$ at all values of $Q^2$
considered here at the leading order of perturbative QCD. We find that the inclusion of NLO terms leads to an enhancement in the results of 
$R_{p}^{\gamma}(x,Q^2)$ (dash-dotted line) as compared to the case of LO in the entire kinematic range of $x$ and $Q^2$, for example, there is
large enhancement at $x<0.4$, that is about $82\%$ at $x=0.1$, $27\%$ at $x=0.3$, however, it becomes small for $x\ge0.4$, for example, $4\%$ at $x=0.6$ 
for $Q^2=2$ GeV$^2$. For $Q^2=8$ GeV$^2$ this enhancement becomes $58\%$ at $x=0.1$, $30\%$ at $x=0.3$ and about $6\%$ at $x=0.6$ which implies that 
the corrections due to NLO terms become small with the increase in $x$. Furthermore, we find that the results for $R_{p}^{\gamma}(x,Q^2)$ 
at NLO with the TMC effect get reduced from the results obtained without the TMC effect. For example,
the reduction in the ratio due to TMC effect is about $55\%$ at $x=0.4$ and $80\%$ at $x=0.6$ for $Q^2=2$ GeV$^2$ which becomes $42\%$ at $x=0.4$
and $73\%$ at $x=0.6$ for $Q^2=8$ GeV$^2$ implying that TMC effect decreases with the increase in $Q^2$. From the figure, it may be noticed that 
the further inclusion of HT effect with the TMC effect at NLO (solid line) leads to an enhancement in the ratio 
$R_{p}^{\gamma}(x,Q^2)$ relative to the results with TMC effect only, however, it becomes small with the increase in $Q^2$. Quantitatively,
the enhancement due to HT effect is about $40\%$ $(30\%)$ at $x=0.4$ and $58\%$ $(50\%)$ at $x=0.6$ for $Q^2=2~(4)$ GeV$^2$ as
compared to the results obtained with TMC effect only. We find that the corrections from NNLO
terms (dash-double dotted line) are small compared to NLO terms (double dash-dotted line) and leads to a slight enhancement in the numerical results.
The results for $R_{p}^{\gamma}(x,Q^2)$ at NNLO with TMC effect only are observed to be in agreement with the results obtained using 
the phenomenological prescription given by Whitlow et al.~\cite{Whitlow:1991uw, Whitlow:1990gk} (solid line with circles), especially in the 
region of low to intermediate $x$. 

The numerical results for $R_{p}^{\gamma Z}(x,Q^2)=\frac{F_{Lp}^{\gamma Z}(x,Q^2)}{2 x F_{1p}^{\gamma Z}(x,Q^2)}$ as a function of $x$ 
for the different values of $Q^2$ are presented in Fig.~\ref{res3} (bottom panel). For the evaluation of $F_{Lp}^{\gamma Z}(x,Q^2)$, we follow the same 
prescription as used for the $F_{Lp}^{\gamma}(x,Q^2)$~\cite{Moch:2004xu}, by taking the appropriate couplings in the case of $\gamma-Z$ interference terms.
It may be pointed out that the qualitative behavior of $R_{p}^{\gamma Z}(x,Q^2)$ with perturbative QCD corrections, and with the nonperturbative
TMC and HT corrections is observed to be similar to the case of $R_{p}^{\gamma}(x,Q^2)$ but are quantitatively different as discussed above.

A comparative study of $R_{p}^{\gamma }(x,Q^2)$ and $R_{p}^{\gamma Z}(x,Q^2)$ is important, as it plays an important role in the dominant term 
of the electron asymmetry given in Eq.~\ref{aee}. In the limit, $R_{p}^{\gamma }(x,Q^2) \to R_{p}^{\gamma Z}(x,Q^2)$, $Y_{1p} \to 1$,
the dominant term in the asymmetry depends only on the ratio $\frac{F_{1p}^{\gamma Z}(x,Q^2)}{F_{1p}^{\gamma}(x,Q^2)}$.
In order to compare $R_{p}^{\gamma }(x,Q^2)$ and $R_{p}^{\gamma Z}(x,Q^2)$, we present the 
results in Fig.~\ref{res4} for $\frac{R_{p}^{\gamma Z}(x,Q^2)}{R_{p}^{\gamma}(x,Q^2)}$ vs $x$ at
$Q^2=2$ GeV$^2$ and $Q^2=8$ GeV$^2$. The numerical
results are evaluated up to NLO, both without and with the corrections due to the TMC and HT effects. We find that even at 
the leading order (LO), the ratio deviates from unity in the low and intermediate region of $x$ ($x\le 0.45$) and could be $6-7\%$ for very small values of $x$.
This deviation increases when NLO terms are included especially in the region of low to intermediate $x$,
however, for high $x (\gtrsim 0.45)$, the ratio approaches unity. We find that the additional contribution due to the inclusion 
of NLO terms (dash-dotted line) relative to LO results (dashed line) diminishes with the increasing $Q^2$. From the figure, it may be noticed that the TMC effect is important
in the present kinematic region of $x$ and $Q^2$, and leads to an enhancement in the region of low to intermediate $x$, while a reduction at high $x$ relative to the 
results obtained at NLO without the TMC effect. 
The effect of HT corrections is found to be comparatively smaller than the TMC effect which results to a reduction in the ratio up to 
the mid region of $x$ while a small enhancement in the high $x$ region. We find that the effect of both the TMC and HT corrections on the ratio
decreases with the increase in $Q^2$.

\begin{figure}
\begin{center}
 \includegraphics[height=8 cm, width=\textwidth]{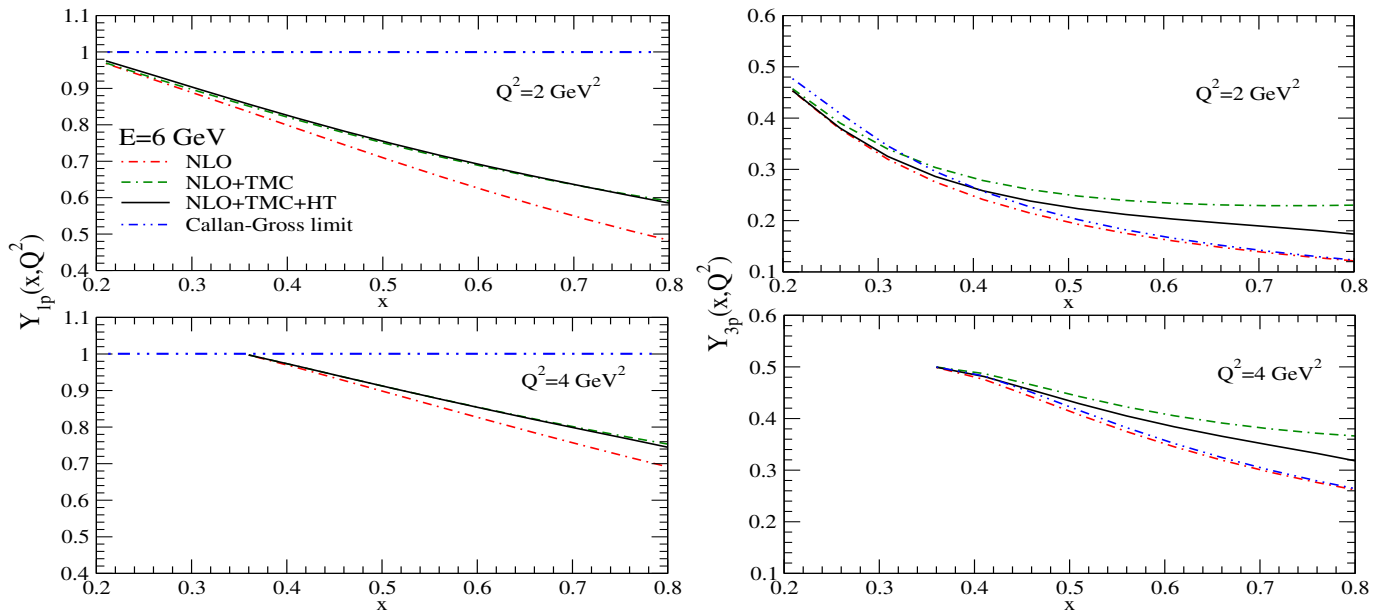}
  \end{center}
   \caption{Results for the ratio $Y_{1p}(x,Q^2)$ (left panel) and $Y_{3p}(x,Q^2)$ (right panel)
     at $Q^2=2$ GeV$^2$ (top panel) and $Q^2=4$ GeV$^2$ (bottom panel). The evaluation is performed at 
 NLO without and with the nonperturbative effects like TMC and HT (as mentioned in the legends of the figure) for $E=6$ GeV. Note: y-axis scales for 
 $Y_{1p}(x,Q^2)$ and $Y_{3p}(x,Q^2)$ are different in the graphs.}
  \label{res5}
\end{figure}

We have obtained the numerical results for the coefficients $Y_{1p}(x,Q^2)$ (left panel)
and $Y_{3p}(x,Q^2)$ (right panel) using the results of $R_{p}^{\gamma Z/\gamma}(x,Q^2)$, and present them in Fig.~\ref{res5}. These coefficients provide 
information about the sensitivity of the vector and axial-vector interactions of the electron to the parity violating asymmetry $A_{PV}^{(e)}(x,Q^2)$ (Eq.~\ref{aee}).
It is noticeable that the contribution of the term with $Y_{3p}(x,Q^2)$ associated with the PV electron asymmetry is smaller than the term with $Y_{1p}(x,Q^2)$ due to $g_V^e$.
The numerical calculations are performed at NLO for $Q^2=2$ GeV$^2$ and $Q^2=4$ GeV$^2$. 
In the Callan-Gross limit ($Q^2\to \infty$), $Y_{1p}(x,Q^2)$ (Eq.~\ref{y13}) is unity, however, in the present kinematic region of $x$ and $Q^2$,
$Y_{1p}(x,Q^2)$ significantly deviates from unity, due to the finite contributions of $R_{p}^{\gamma}(x,Q^2)$ and $R_{p}^{\gamma Z}(x,Q^2)$. 
This deviation from unity is smaller at low $x$, but increases with the increase in $x$. For example, in the region of $0.5\le x \le 0.8$
at $Q^2=2$ GeV$^2$, the change in $Y_{1p}(x,Q^2)$ from unity due to $R_{p}^{\gamma Z/\gamma}(x,Q^2)$ evaluated at NLO with TMC and HT effects (solid line)
is about 25-40\%. At $Q^2=4$ GeV$^2$, this reduces to 10-25\% for the same $x$ range. 
In contrast to $Y_{1p}(x,Q^2)$, the coefficient $Y_{3p}(x,Q^2)$, which contributes to the
axial-vector part in electron asymmetry, depends only on $R_{p}^{\gamma}(x,Q^2)$. The numerical results with TMC and HT corrections show significant
deviation from the Callan-Gross limit, $f(y)$ defined before Eq.~\ref{aes0}, which ranges $0.477\le f(y) \le 0.12$ for $Q^2=2$ GeV$^2$ 
and $0.5\le f(y) \le 0.26$ for $Q^2=4$ GeV$^2$, in the kinematic region of $0.21\le x \le 0.8$ and $0.36\le x \le 0.8$, respectively. 
Specifically, the variation in $Y_{3p}(x,Q^2)$ from the Callan-Gross limit is about 40\% at $Q^2=2$ GeV$^2$ and $20\%$ at $Q^2=4$ GeV$^2$ for $x=0.8$.
Furthermore, we find that the TMC effect is significant in $Y_{1p}(x,Q^2)$, whereas the HT corrections are relatively small.

\subsection{Differential cross sections and parity violating electron asymmetry}

\begin{figure}
 \includegraphics[height=8 cm, width=\textwidth]{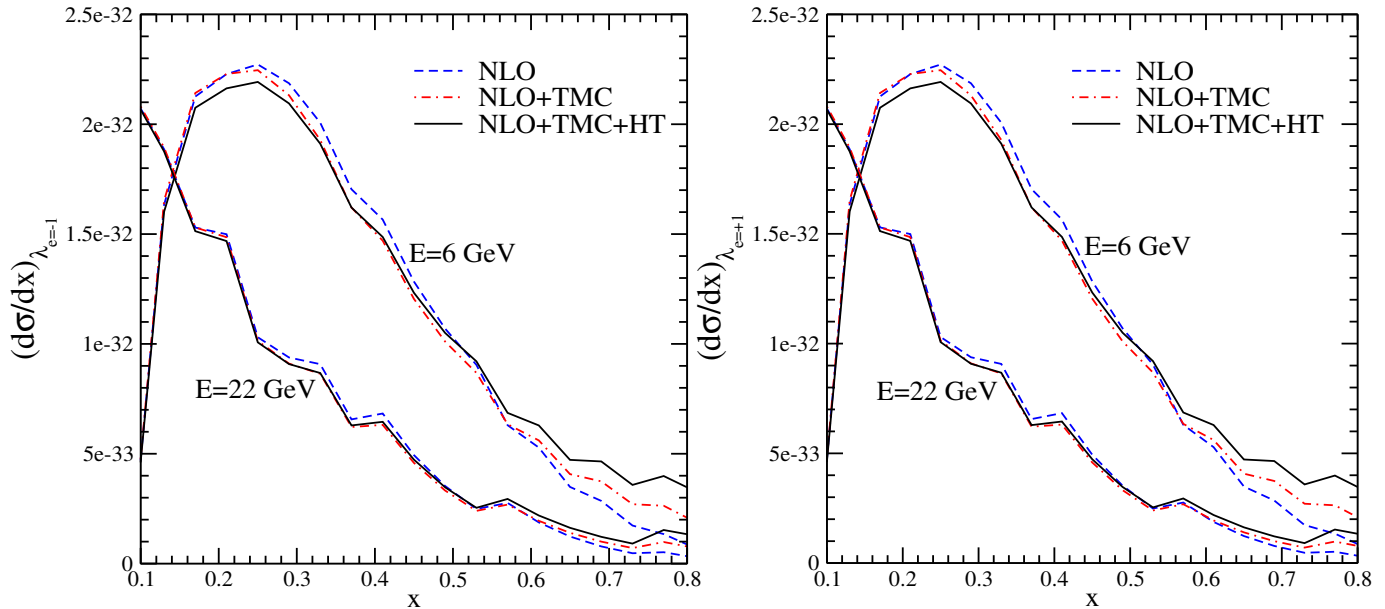}
 \caption{Polarized $\vec {e} p$ electroweak single differential cross sections $\Big(\frac{d\sigma^{ep}}{dx}\Big)_{\lambda_e=-1}$ (left panel)
  and $\Big(\frac{d\sigma^{ep}}{dx}\Big)_{\lambda_e=+1}$ (right panel) vs $x$ for $E=6$ GeV and $E=22$ GeV beam energies. The results 
 are evaluated at NLO without and with the TMC and higher twist effects incorporating the contributions from $\gamma-$exchange
 and $\gamma Z$ interference channels. No cut on the center of mass energy is applied here. }
   \label{res8}
\end{figure}

After evaluating the proton structure functions with higher order perturbative and nonperturbative corrections, we use them to
evaluate the differential scattering cross sections (Eq.~\ref{xsec_spin}). In Fig.~\ref{res8}, we present the numerical results 
for the single differential scattering cross section $\frac{d\sigma^{ep}}{dx}$ by performing the integration 
over the inelasticity $y (\in[0,1])$ in Eq.~\ref{xsec_spin} for both the negative ($\lambda_e=-1$: left panel) and 
positive ($\lambda_e=+1$: right panel) helicity states of the projectile beam of polarized electrons with energy $E=6$ GeV and $E=22$ GeV. These results are evaluated at 
NLO without and with the TMC and HT corrections. It may be observed from the figure that the differential cross sections corresponding to 
the higher beam energy have a sharp peak in the very low region of $x$, while for the lower beam energy it has a broader spectrum 
with the shift in peak towards the intermediate region of $x$. The qualitative behavior of the 
QCD corrections is similar for both $\Big(\frac{d\sigma^{ep}}{dx}\Big)_{\lambda_e=-1}$ and $\Big(\frac{d\sigma^{ep}}{dx}\Big)_{\lambda_e=+1}$, however, 
quantitatively they are different, leading to very small corrections in the electron beam asymmetry $A_{PV}^{(e)}(x,Q^2)$. The inclusion of TMC and HT effects
is not significant except at very high $x$, i.e., around $x \approx 0.7$ and above.

\begin{figure}
 \includegraphics[height=8 cm, width=\textwidth]{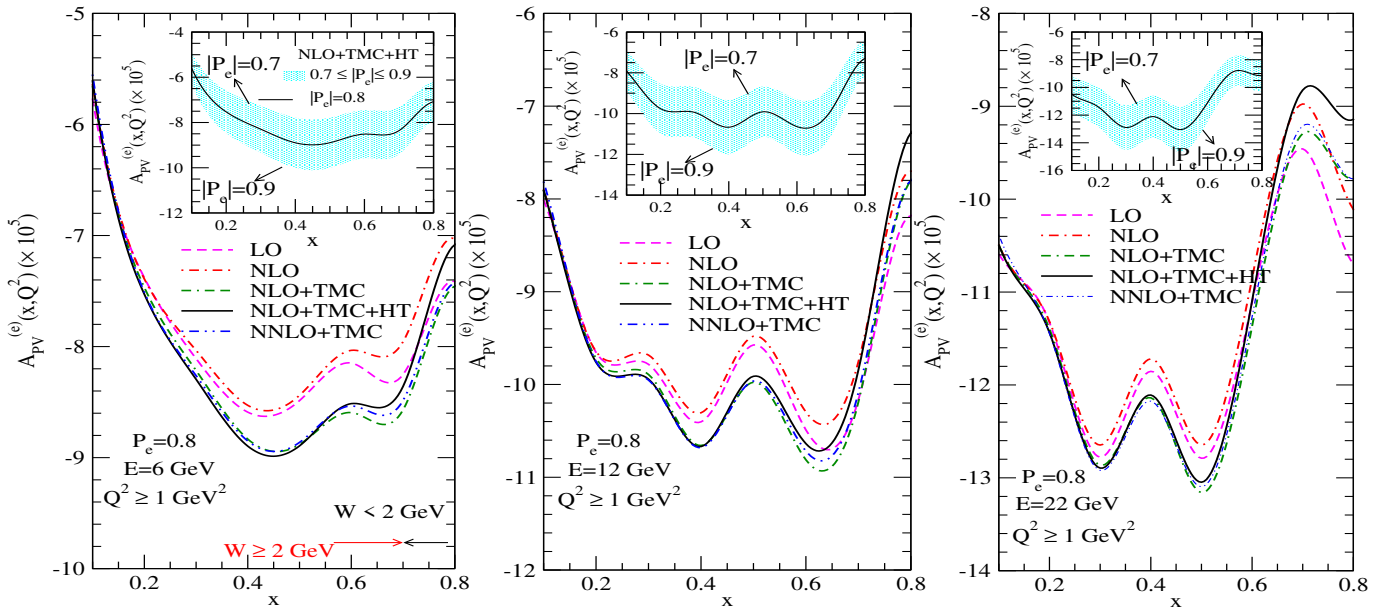}
 \caption{Parity violating single electron asymmetry $A_{PV}^{e}(x,Q^2)$ vs $x$ for different beam energies with 80\% longitudinal polarization,
 i.e., $P_e=0.8$. The results 
 are evaluated at LO without TMC and HT corrections, at NLO without and with the TMC and higher twist effects as well as at NNLO with TMC effect. No cut on center of mass energy is applied here. In the 
 inset $A_{PV}^{e}(x,Q^2)$ at NLO with TMC and HT effects is shown for different values of longitudinal beam polarization,
 $0.7\le \mid P_e\mid\le 0.9$ (cross pattern band) at $E=6$ GeV (left panel), $E=12$ GeV (middle panel) and $E=22$ GeV (right panel).}
   \label{res10}
\end{figure}
The electron spin asymmetry $A_{PV}^{(e)}(x,Q^2)$ vs $x$ is evaluated by using Eq.~\ref{ase} for different beam energies 
viz. $E=6$ GeV, 12 GeV and 22 GeV in the kinematic region of $Q^2\ge 1$ GeV$^2$, and the results are shown in Fig.~\ref{res10}. We have chosen 
the beam polarization to be $|P_e|=0.8$, keeping in mind the future experiments like EIC, JLab upgrade, and the EicC. The present numerical
calculations are performed by evolving the parton densities up to NNLO and by incorporating the TMC and HT corrections to explicitly observe their 
effect on the electron spin asymmetry. 
\begin{itemize}
 \item It may be noted that $A_{PV}^{(e)}(x,Q^2)$ is negative and its absolute value increases with increasing beam energy for all values of $x$  considered here. We find that the
 corrections due to the next-to-leading order terms in the perturbative expansion are quite small
 in the entire kinematic region of $x$. For example, the enhancement in the results for $A_{PV}^{e}(x,Q^2)$ evaluated at NLO from 
the results at LO is about 1\% at $x=0.6$ and 4\% at $x=0.8$ for $E=6$ GeV, while for $E=22$ GeV it becomes 2\% at $x=0.6$ and 6\% at $x=0.8$.
We also obtain the numerical results at NNLO and find that the corrections due to NNLO terms are still smaller as compared to the corrections due to the NLO terms
in the present kinematic region of $x$ and $E$ (not shown here explicitly). 

 \item Moreover, the effect of target mass corrections are also small except in the region of high $x$ in the case of 
 beam energies around 6 GeV, and it decreases with increasing beam energy. For example, at $x=0.4$ the change 
in $A_{PV}^{(e)}(x,Q^2)$ due to the inclusion of 
TMC effect at NLO is about 2-3\% for the beam energies considered here, while at $x=0.6$ it becomes 6\% for $E=6$ GeV and 3\% for $E=22$ GeV. 

 \item The higher twist effect is found to be significant in the region of high $x$ and increases with increasing beam energy $E$. 
Quantitatively, at $x=0.8$, the change in $A_{PV}^{(e)}(x,Q^2)$ due to the inclusion of HT corrections alongside TMC effect
at NLO is approximately 5\% and 7\% for $E=6$ GeV and 22 GeV, respectively, relative to the results obtained with the TMC effect alone at NLO. 

 \item It is important to point out that the results for 
$A_{PV}^{(e)}(x,Q^2)$ obtained at NLO with TMC and HT effects (solid line) are similar with the results at NNLO with
TMC effect only (dash-double dotted line) in the region of low to intermediate $x$, while for $x > 0.6$ there is difference between them
which is quite small.

\item The effect of all the corrections taken together is rather small across the entire $x$ region, except for the case of high $x$ specifically in the region of low 
beam energies. 
%
 We have also studied the variation in $A_{PV}^{(e)}(x,Q^2)$ for different values of beam polarization vector $0.7 \le \mid P_e \mid \le 0.9$ 
and it is shown by a dotted-filled pattern band in the inset of the figure~\ref{res10}. These results are obtained at NLO by taking into account the TMC and HT effects.
\end{itemize}

\subsection{Electroweak structure functions and $d(x)/u(x)$ ratio}
\begin{figure}
 \includegraphics[height=8 cm, width=\textwidth]{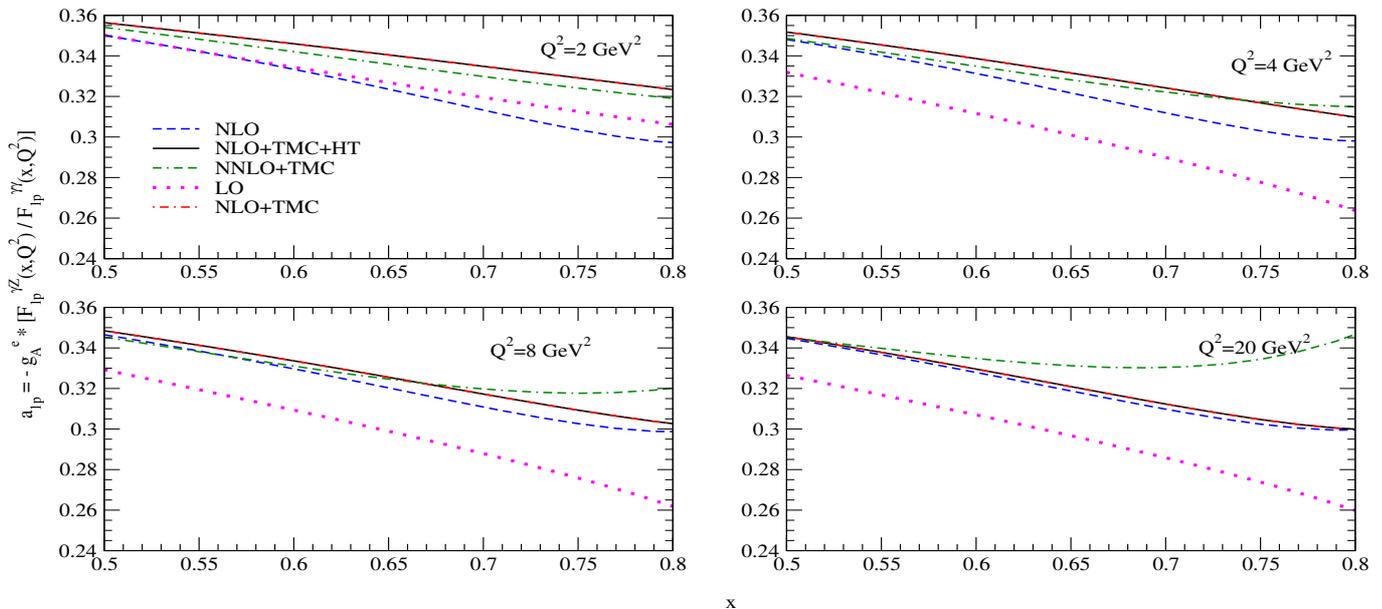}
     \caption{Results for the ratio $a_{1p}(x,Q^2)=-g_A^e\;\frac{ F_{1p}^{\gamma Z}(x,Q^2)}{F_{1p}^{\gamma}(x,Q^2)}$ vs $x$
     at different values of $Q^2$. The evaluation is performed at 
 NLO and NNLO without and with the nonperturbative effects like TMC and HT (as mentioned in the legends of the figure).}
  \label{res6}
\end{figure}

The determination of $d(x)/u(x)$ ratio is a topic of interest for both the experimenters as well as theorists as it 
directly reflects how momentum is shared between these quarks as a function of Bjorken $x$. In recent years, the study of the parity violating
DIS off proton target in the region of large $x$ is considered to be a important tool to obtain information about $d(x)/u(x)$ ratio.
Traditionally, the $d(x)/u(x)$ ratio has been determined from the electromagnetic EMC measurements of $\frac{F_{2A}(x,Q^2)}{F_{2p}(x,Q^2)}$
on nuclei like $A=2$($^2_1 D$), $3$($^3_1 H$, $^3_2 He$)~\cite{Cocuzza:2026vey}. However, measurements of $F_{2A}(x,Q^2)$  
are beset with the uncertainties 
due to the presence of nuclear effects and the lack of clear demarcation between the shallow inelastic scattering and DIS transition region due to the 
ambiguities in defining precise cuts on the center of mass energy $W$ and the four momentum transfer square $Q^2$~\cite{SajjadAthar:2020nvy}. In the case of proton
target, there are no nuclear medium corrections, however, uncertainties related to the proper definition of the boundary of the transition region still
present. Owing to the absence of nuclear effects, studies of parity violating electron asymmetry for proton targets have been 
proposed as a cleaner method to determine the $d(x)/u(x)$ ratio and to investigate possible violations of isospin asymmetry using future data
in the region of high $x$ in which sea quarks contribution can be neglected~\cite{Brady:2011uy, Cocuzza:2026vey}. 

Since $a_{3p}(x,Q^2)=-g_V^e\;\frac{ F_{3p}^{\gamma Z}(x,Q^2)}{F_{1p}^{\gamma}(x,Q^2)}$ is very small, therefore, parity violating electron asymmetry $A_{PV}^{(e)}(x,Q^2)$
depends explicitly on the first term in Eq.~\ref{aee}. 
For example, we have found that at the leading
order of perturbative QCD, the contribution of the term associated with $a_{3p}(x,Q^2)$ to the parity violating electron asymmetry 
$A_{PV}^{(e)}(x,Q^2)$ (not shown here explicitly) is at the most 8-10\% in the present kinematic region of $x$ for $E=6$ GeV.
In the region of high $x$, $R_p^\gamma(x,Q^2)=R_p^{\gamma Z}(x,Q^2)$ as shown in Fig.~\ref{res4}, therefore, 
$Y_{1p}(x,Q^2) \to 1$(using Eq.~\ref{y13}) and the asymmetry(Eq.~\ref{aee}) depends dominantly on $a_{1p}(x,Q^2)$. Moreover, in the high $x$ region, 
the contribution from the 
sea quarks can be neglected. In this situation, $a_{1p}(x,Q^2)$ is given by
\begin{eqnarray}
 a_{1p}(x)&=&\frac{\sum_i\;C_{1i} e_i q_i(x)}{\sum_i e_i^2 q_i(x)},\hspace{3 mm}
\end{eqnarray} 
After simplification, this ratio is expressed in terms of the effective coupling constants ($C_{iq};~i=1-2$) and quark densities as:
 \begin{eqnarray}\label{a1p_app}
  a_{1p}(x,Q^2)&=&\frac{6 C_{1u}-3 C_{1d}\;\frac{d(x,Q^2)}{u(x,Q^2)}}{4+\frac{d(x,Q^2)}{u(x,Q^2)}}\;\;\;\;
 \end{eqnarray}
It may be noted that $a_{1p}(x,Q^2)$ provides direct information about the $d(x)/u(x)$ ratio in the kinematic region of large $x$ to which the 
electron asymmetry $A_{PV}^{(e)}(x,Q^2)$ is sensitive. Since in this kinematic region of $x$, nonperturbative QCD corrections are observed to 
be important in the proton structure functions, hence, the theoretical study of $a_{1p}(x,Q^2)$ in the presence of these corrections
is relevant for understanding the differences in quark flavors and possible charge symmetry violation. 

 Therefore, in Fig.~\ref{res6}, we present the numerical results for
 $a_{1p}(x,Q^2)$ vs $x$ ($0.5\le x \le 0.8$) at the different values of $Q^2$. It may be noticed from this figure that at the leading order
 the ratio $a_{1p}(x,Q^2)$ decreases with the increase in $x$ and $Q^2$. However, when higher order perturbative and nonperturbative corrections 
 are incorporated the ratio gets enhanced relative to the results at LO.
For example, due to the inclusion of NLO terms (dashed line) we find an enhancement of 
 $<1\%$ at $x=0.6$ and $3\%$ at $x=0.8$ for $Q^2=2$ GeV$^2$ relative to the results at LO, and it becomes 6\% and 14\% for $Q^2=8$ GeV$^2$. There is 
 further enhancement in $a_{1p}(x,Q^2)$ when the contribution of NNLO terms is taken into account which is about 
 $2\%$ ($<1\%$) at $x=0.6$ and $8\%$ ($13\%$) at $x=0.8$ for $Q^2=2$ (8) GeV$^2$ from the results evaluated at NLO. 
Inclusion of TMC effect leads to an enhancement of about $4\%$ (2$\%$) at $x=0.6$ and $8\%$ ($3\%$) at $x=0.8$ for $Q^2=2 (4)$ GeV$^2$, whereas, the
effect of HT corrections on $a_{1p}(x,Q^2)$ is found to be almost negligible. 


Using Eq.~\ref{a1p_app} and taking $a_{1p}(x,Q^2)$ as an input,
we have evaluated the effect of QCD corrections on the $d(x)/u(x)$ ratio and compared it with the results at LO. In the case of 
perturbative corrections, we find that the inclusion of NLO terms for $Q^2=2$ GeV$^2$,
there is a reduction in the $d(x)/u(x)$ ratio of about 3\% at $x=0.6$, 23\% at $x=0.7$ and $48\%$ at $x=0.8$ from the
results at LO. Whereas, there is a further reduction in the $d(x)/u(x)$ ratio due to the inclusion of NNLO terms which 
is about $8\%$, $\sim 19\%$ and $15\%$ from the 
results evaluated at NLO for the same values of $x$. When the target mass correction effect is taken into account at NLO, $d(x)/u(x)$ ratio gets enhanced 
relative to the case without including it, for example, 25\% at $x=0.6$, 52\% at $x=0.7$ and $73\%$ at $x=0.8$. The effect of HT corrections is 
negligible.

\section{Summary and conclusion}\label{summ}
 In this work, we have studied the effects of perturbative and nonperturbative QCD corrections to the QPM in the evaluation of the 
 proton structure functions in the electromagnetic as well as in the weak-electromagnetic interference sectors. We have also studied the impact of these 
 corrections on the Callan-Gross relation~\cite{Callan:1969uq} by calculating the values of the ratios 
 $r^{\gamma Z/\gamma}(x,Q^2)=\frac{F_{2p}^{\gamma Z/\gamma}(x,Q^2)}{2 x F_{1p}^{\gamma Z/\gamma}(x,Q^2)}$ and 
 $R_{p}^{\gamma Z/\gamma}(x,Q^2)=\frac{F_{Lp}^{\gamma Z/\gamma}(x,Q^2)}{2 x F_{1p}^{\gamma Z/\gamma}(x,Q^2)}$. The differential scattering cross 
 sections $\frac{d\sigma}{dx}$ and the parity violating electron beam asymmetry $A_{PV}^{(e)}(x,Q^2)$ have also been studied. The higher order perturbative
 corrections are calculated up to NNLO using MMHT PDFs in the 3-flavor $\overline{\textrm{MS}}$ scheme. The 
nonperturbative effects, namely, the target mass corrections and the higher twist corrections have been incorporated. The effect of 
these QCD corrections on the ratio $d(x)/u(x)$ in the region of high $x$ has also been discussed. 

We conclude that:

 \begin{itemize}
  \item The effect of higher order perturbative corrections are quantitatively significant on the differential cross sections and the parity violating electron asymmetry only in the 
region of high $x$, i.e., $x\ge 0.6$.

\item The TMC effect is found to be significant in the differential scattering cross sections as well as 
the parity violating electron asymmetry in the region of intermediate-to-high $x ~(\gtrsim 0.4)$ while the higher twist corrections are observed to be important in the 
kinematic region of high $x~(\gtrsim 0.6)$ for the projectile beam energies considered in the present work. 

\item The violation of the Callan-Gross relation due to nonperturbative QCD corrections is significant in the region of intermediate-to-high $x$ 
for all the values of $Q^2$ in the range of $2\le Q^2 \le 20$ GeV$^2$ considered in this work. The violation of Callan-Gross relation is found to be quantitatively different
in the electromagnetic and weak-electromagnetic structure functions of the proton.

\item  The present study of $a_{1p}(x,Q^2)$ with QCD corrections may be helpful in understanding the behavior of $d(x)/u(x)$ ratio 
as well as in interpreting the relevant experimental data, especially, from the upcoming experiments at JLab.

 \end{itemize}
 Thus to summarize, the higher order QCD corrections on the proton structure functions are quite significant for both the $\gamma-$exchange and $\gamma Z-$exchange channels
in the present kinematic region of $0.\le x \le 0.8$ and $2\le Q^2\le 20$ GeV$^2$ relevant to the ongoing and upcoming experiments at JLab, EIC and EicC, and 
a better theoretical understanding is required in the region of high $x$, where nonperturbative effects are important.

\section*{Acknowledgment}
We thank D. Indumathi for valuable feedback on the manuscript.
F. Zaidi is thankful to Council of Scientific \& Industrial Research, Govt. of India for providing Senior Research Associate-ship (SRA) under the Scientist's Pool Scheme, file no. 13(9240-A)2023-POOL and to 
the Department of Physics, Aligarh Muslim University, Aligarh for providing the necessary facilities to pursue this research work.
M. S. A. is thankful to the Department of Science and Technology (DST), Government of India for providing 
financial assistance under Grant No. SR/MF/PS-01/2016-AMU/G.


%
%
%
%
%
%
%
%
%
%
%
%
%
%
%
%
%
%
%
%
%
%
%
%
%
%
%
%
%
%
%
%
%
%
%
%
%
%
%
%
%
%
%
%
%
%
%
%
%
%
%
%
%
%
%
%
%
%
%
%

\end{document}